\documentclass[12pt]{article}

\usepackage{lineno}
\usepackage{amsmath,amssymb, amsbsy}
\usepackage{graphicx}
\usepackage{endfloat}
\usepackage{bbold}

\usepackage{textcomp}
\usepackage{abstract}
\usepackage[normalem]{ulem}
\usepackage[default]{jasa_harvard}
\usepackage{JASA_manu}
\usepackage[letterpaper,textwidth=6in, textheight=8.5in]{geometry}

\usepackage{xcolor}
\usepackage{color}

\newcommand{\Var}{\operatorname{Var}}
\newcommand{\Median}{\operatorname{Median}}
\newcommand{\Gammad}{\operatorname{Gamma}}

\newtheorem{proposition}{Proposition}
\newtheorem{lemma}[proposition]{Lemma}

\begin{document}

\title{A Bayesian Model of NMR Spectra for the Deconvolution and Quantification of Metabolites in Complex Biological Mixtures}
\author{\textsuperscript{\textborn}William Astle, \textsuperscript{\textleaf}Maria De Iorio, \textsuperscript{\textdagger}Sylvia Richardson,\\
\textsuperscript{\textborn\textborn}David Stephens and \textsuperscript{\textdaggerdbl}Timothy Ebbels.\\ \\
\textsuperscript{\textborn}Department of Epidemiology, Biostatistics and Occupational Health,\\
\textsuperscript{\textborn\textborn}Department of Mathematics and Statistics,\\McGill University, Montreal, Canada.\\\\
\textsuperscript{\textleaf}Department of Statistical Science, University College, London,\\
United Kingdom. \\
\\
\textsuperscript{\textdagger}Department of Epidemiology and Biostatistics,\\
\textsuperscript{\textdaggerdbl}Section of Biomolecular Medicine,  Department of Surgery and Cancer,
\\Faculty of Medicine, Imperial College, London. United Kingdom.}

\clearpage
\maketitle
\thispagestyle{empty}
\newpage
\thispagestyle{empty}
\linenumbers

\textbf{Authors' footnote}\\
William Astle (E-mail: william.astle@mcgill.ca) is a Postdoctoral Fellow in the Department of Epidemiology, Biostatistics and Occupational Health, McGill University, Purvis Hall, 1020 Ave.~des Pins Ouest, Montreal, QC, H3A 1A2, Canada. Maria De Iorio (E-mail:
m.deiorio@ucl.ac.uk) is a Reader in Statistical Science in the Department of Statistical Science, University College, Gower Street,
London, WC1E 6BT, UK. Sylvia Richardson (E-mail: sylvia.richardson@ic.ac.uk) is Professor of Biostatistics in the Department of Epidemiology and Biostatistics, Imperial College, St.
Mary's Hospital Campus, London, W2 1PG, UK. David Stephens
(E-mail: d.stephens@math.mcgill.ca) is Professor of Statistics in the
Department of Mathematics and Statistics, McGill University, Montreal, QC,
Canada, H3A 2K6. Timothy Ebbels (E-mail: t.ebbels@ic.ac.uk) is a Senior Lecturer
in Computational Bioinformatics in the Section of Biomolecular Medicine,
Department of Surgery and Cancer, South Kensington Campus, Imperial College,
London. SW7 2AZ. This work was supported by UK BBSRC grant BB/E020372/1.  Dr. Stephens is supported by a Discovery Grant from the Natural Sciences and Engineering Council of Canada. Dr. Astle's present position is supported by a Team Grant from the Fonds de recherche du Qu\'{e}bec - Nature et technologies. The authors thank Jake Bundy for providing the yeast dataset and for help interpreting deconvolutions of the NMR spectra, Jie Hao for programming a C++ implementation and David Balding and Ernest Turro for comments on a draft manuscript.

\newpage
\begin{abstract}
\thispagestyle{empty} 

Nuclear Magnetic Resonance (NMR) spectra are widely used in metabolomics to obtain profiles of metabolites dissolved in biofluids such as cell supernatants. Methods for estimating metabolite concentrations from these spectra are presently confined to manual peak fitting and to binning procedures for integrating resonance peaks.  Extensive information on the patterns of spectral resonance generated by human metabolites is now available in online databases. By incorporating this information into a Bayesian model we can deconvolve resonance peaks from a spectrum and obtain explicit concentration estimates for the corresponding metabolites. Spectral resonances that cannot be deconvolved in this way may also be of scientific interest so we model them jointly using wavelets.

We describe a Markov chain Monte Carlo algorithm which allows us to sample from the joint posterior distribution of the model parameters, using specifically designed block updates to improve mixing. The strong prior on resonance patterns allows the algorithm to identify peaks corresponding to particular metabolites automatically, eliminating the need for manual peak assignment.

We assess our method for peak alignment and concentration estimation. Except in cases when the target resonance signal is very weak, alignment is unbiased and precise. We compare the Bayesian concentration estimates to those obtained from a conventional numerical integration method and find that our point estimates have sixfold lower mean squared error.

Finally, we apply our method to a spectral dataset taken from an investigation of the metabolic response of yeast to recombinant protein expression. We estimate the concentrations of $26$ metabolites and compare to manual quantification by five expert spectroscopists.  We discuss the reason for discrepancies and the robustness of our methods concentration
estimates.
\end{abstract}
\vspace*{.3in}

\noindent\textsc{Keywords}: {metabolomics, concentration estimation, prior information, multi component model, block updates.}

\newpage

\setcounter{page}{1}

\section{Introduction}

Metabolomics (also known as \textit{metabonomics} or sometimes \textit{metabolic profiling}) is a scientific discipline concerned with the quantitative study of metabolites, the small molecules that participate in metabolic reactions. Research in this field is expanding rapidly, with applications in many areas of biology and medicine including cancer (e.g. \citeasnoun**{PubMed_11830521}), toxicology (e.g. \citeasnoun**{PubMed_12662897}), organism classification (e.g. \citeasnoun**{PubMed_12067738}), genetics (e.g. \citeasnoun**{PubMed_20037589}), biochemistry (e.g. \citeasnoun**{PubMed_11135551}), epidemiology (e.g. \citeasnoun**{PubMed_18425110}) and disease diagnostics (e.g. \citeasnoun**{brindle2002rapid}). 

Almost all experiments in metabolomics rely on measurements of the abundances of metabolites in complex biological mixtures, often biofluid or tissue samples. One of the most extensively used techniques for obtaining such quantitative information is proton nuclear magnetic resonance (\textsuperscript{1}H NMR) spectroscopy. Metabolites generate characteristic resonance signatures in \textsuperscript{1}H NMR spectra and each signature appears with intensity proportional to the concentration of the corresponding metabolite in the biological mixture.

Specialized models and tools are needed to draw inferences from \textsuperscript{1}H NMR spectroscopic datasets, which are large and heavily structured. At present there is no statistical method for analyzing metabolomic NMR spectra reflecting the data generating mechanisms and the extensive prior knowledge available, e.g. on the form of metabolite NMR signatures.  In this paper, we describe novel Bayesian approaches for the analysis of \textsuperscript{1}H NMR data from complex biological mixtures. Specifically, we develop new models reflecting the data generation mechanisms and our prior knowledge, using a combination of parametric functions and wavelets. We introduce a computational strategy based on Markov chain Monte Carlo (MCMC), including novel block updates to overcome strong posterior correlation between the parametric functions and the wavelets. We demonstrate the utility of the approach with simulations and analyses of data from a yeast metabolomics experiment.

\subsection{NMR spectroscopy}\label{sec:nmrspec}

\begin{figure}
\includegraphics[width=0.99\textwidth]{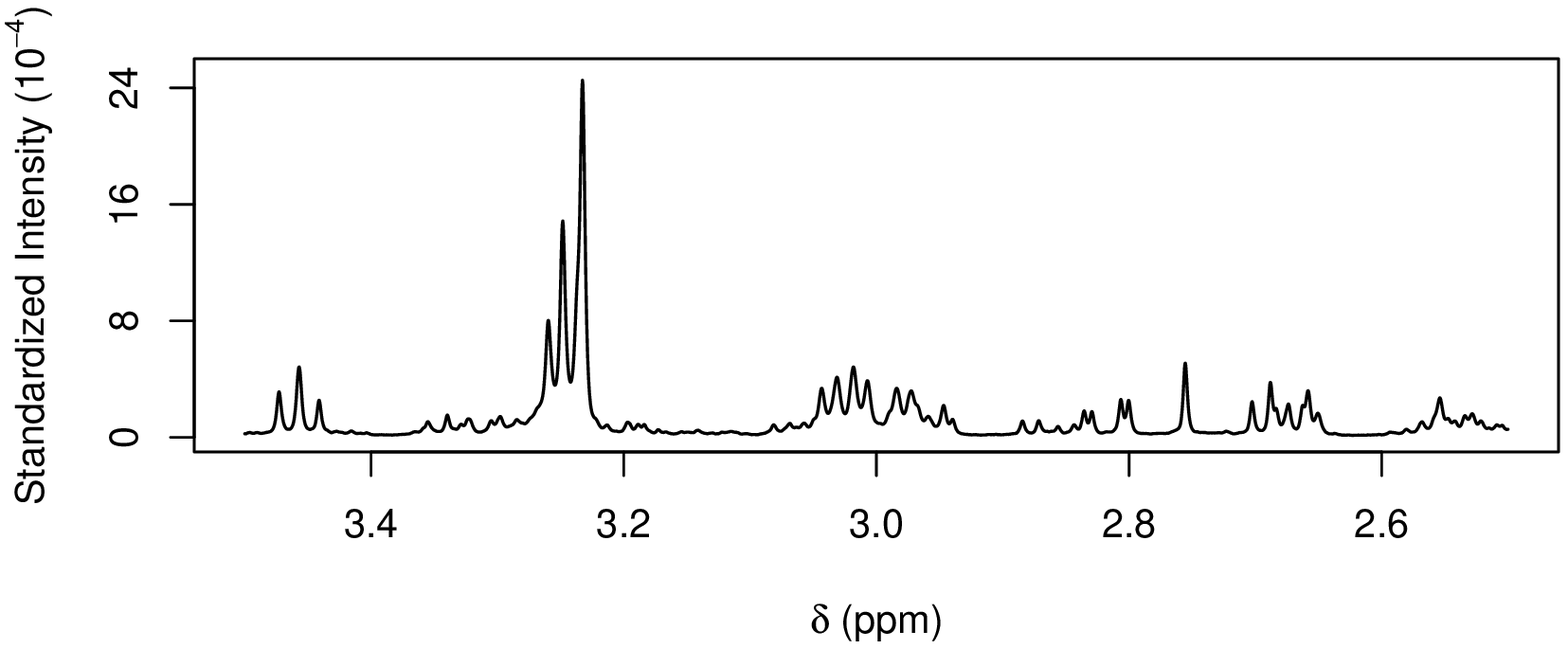}
\caption[Section from an NMR spectrum from a yeast experiment.]{\label{fig:int:exam_spec} A section from an NMR spectrum from an experiment investigating protein expression in yeast. The $x$-axis measures chemical shift in parts per million (ppm). The $y$-axis measures relative resonance intensity.}
\end{figure}

An NMR spectrum consists of a series of measurements of resonance intensity usually taken on a grid of equally spaced frequencies.  Figure \ref{fig:int:exam_spec} is a representative section (of about one tenth) of an NMR spectrum taken from an experiment into the metabolomic response of yeast to recombinant protein expression. The $x$-axis of the spectrum corresponds to resonant frequency and the $y$-axis to resonance intensity.

The spectrum is a collection of convolved peaks with different horizontal positions and vertical scalings, each of which has the form of a Lorentzian curve. The zero centered, standardized Lorentzian function takes the form

\begin{equation}\label{eqn:lor}
l_\gamma(x)=\frac{2}{\pi} \frac{\gamma}{4x^2+\gamma^2}.
\end{equation}
This is the pdf of a Cauchy distribution with scale parameter $\gamma/2$; in
spectroscopy $\gamma$ is called the \textit{peak-width at half-height} (or
sometimes the \textit{linewidth}).

Each spectral peak corresponds to magnetic nuclei resonating at a particular frequency in the biological mixture. This frequency determines the displacement of the peak on the $x$-axis, which is known as its \textit{chemical shift} and is measured in parts per million (ppm) of the resonant frequency of a standard peak. It is conventional in NMR spectroscopy to use $\delta$ to denote chemical shift and for the $\delta$-axis to increase from right to left. \textsuperscript{1}H NMR only detects the resonance of hydrogen nuclei and a typical \textsuperscript{1}H NMR spectrum has a range of about 0ppm-10ppm.

The resonant frequencies of a magnetic nucleus are largely determined by its molecular environment, that is the chemical structure of the molecule in which it is embedded and the configuration of its chemical bonding within the molecule. Consequently, every metabolite has a characteristic molecular \textsuperscript{1}H NMR \textit{signature}, a convolution of Lorentzian peaks that appear in specific positions in \textsuperscript{1}H NMR spectra. These are the peaks observed in the \textsuperscript{1}H NMR spectrum of a pure solution of the metabolite. The peaks of a signature can have quite different chemical shifts (when they are generated by protons with different bonding configurations) and so appear widely separated in a spectrum.

\begin{figure}
\begin{center}
\includegraphics[width=0.99\textwidth,clip]{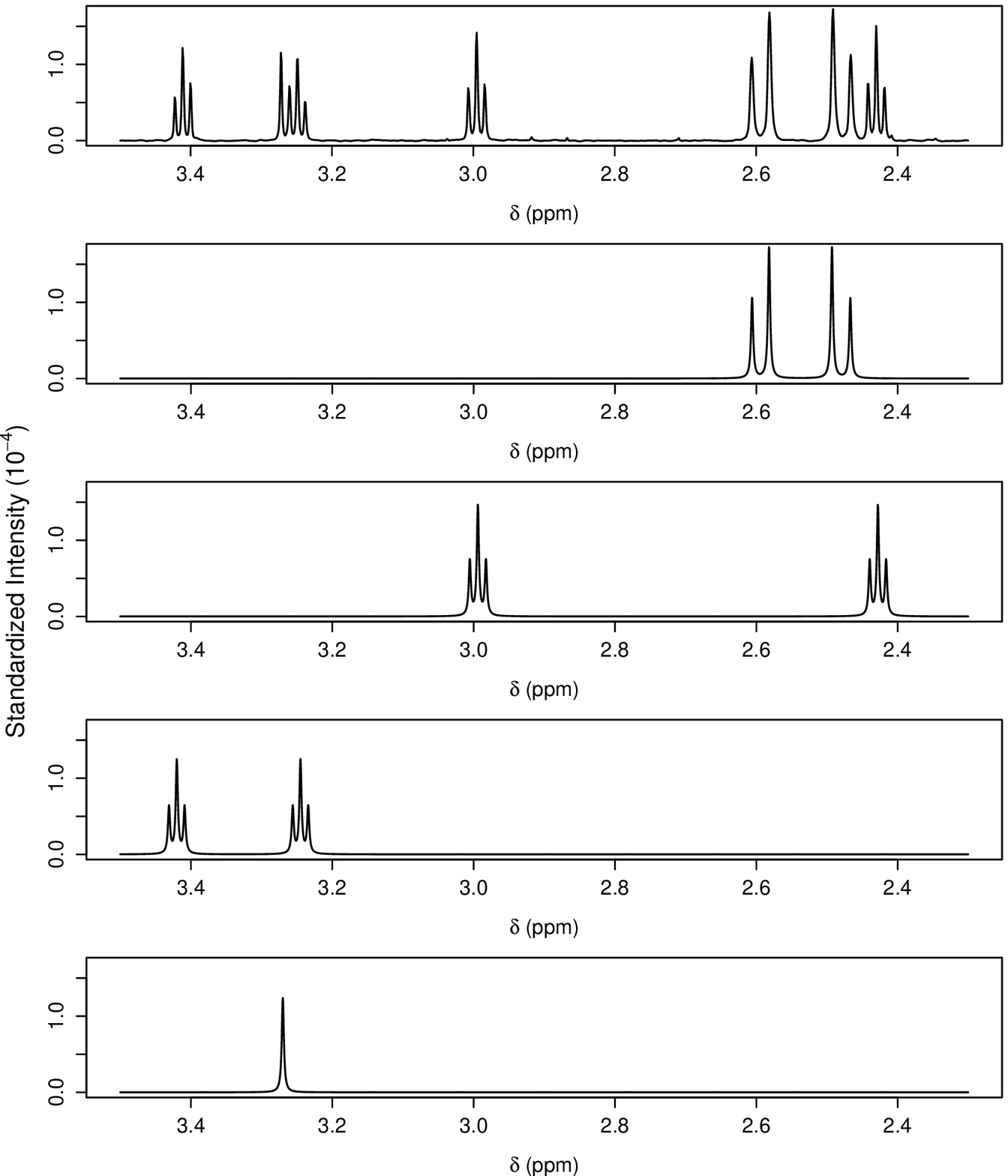}
\end{center}
\caption[Decomposition of a metabolite NMR spectrum]{\label{fig:conv}An \textsuperscript{1}H NMR spectrum (\textit{top panel}) with the principal resonance signals deconvolved into the metabolite NMR signatures (\textit{lower panels}) of (in descending order by panel) citric acid, $2$-oxoglutaric acid, taurine and trimethylamine N-oxide.}
\end{figure}

Depending on its molecular environment, a proton may have more than one resonant frequency and when this happens the frequencies are usually very similar. Consequently, the peaks generated appear in a spectrum as a juxtaposition called a \textit{multiplet}. The shape of a multiplet (number of peaks, their separations and relative heights) can be used to identify the corresponding metabolite. Figure \ref{fig:conv} shows an \textsuperscript{1}H NMR spectrum (top panel) and the resonance signatures of the four metabolites contributing the principal resonance signals (lower panels), with characteristic peak locations and multiplet shapes.

The intensity of a nuclear magnetic signal is proportional to the number of magnetically equivalent nuclei generating the resonance in the biological mixture. Consequently, every resonance peak (and therefore also every metabolite resonance signature) scales vertically in a spectrum in proportion to the molecular abundance of the corresponding compound in the mixture.

\subsection{Specific challenges of NMR in metabolomics}\label{sec:challanges}

Biofluids and tissue samples usually contain thousands of metabolites. However, NMR is relatively insensitive, so that ordinarily a spectrum contains quantitative information on just a few hundred of the most abundant compounds. These compounds can generate hundreds of resonance peaks in a spectrum, many of which overlap.

To quantify a collection of metabolites using NMR, at least one resonance peak generated by each compound must be identified in the spectrum and deconvolved. (To reduce uncertainty, it is desirable to identify as many peaks as possible for each compound.) Estimates of the relative concentrations of the metabolites in the biological sample can be made by comparing the areas under the deconvolved resonance peaks. (Estimates of absolute concentration require a reference compound).

The peak identification step (\textit{assignment}) is complicated by fluctuations in peak positions between spectra, caused by uncontrollable differences in experimental conditions and differences in the chemical properties of the biological samples, such as the pH and ionic strength. When this \textit{positional noise} combines with peak overlap, assignment can become very hard indeed.

\begin{figure}
\begin{center}
\includegraphics[width=0.99\textwidth]{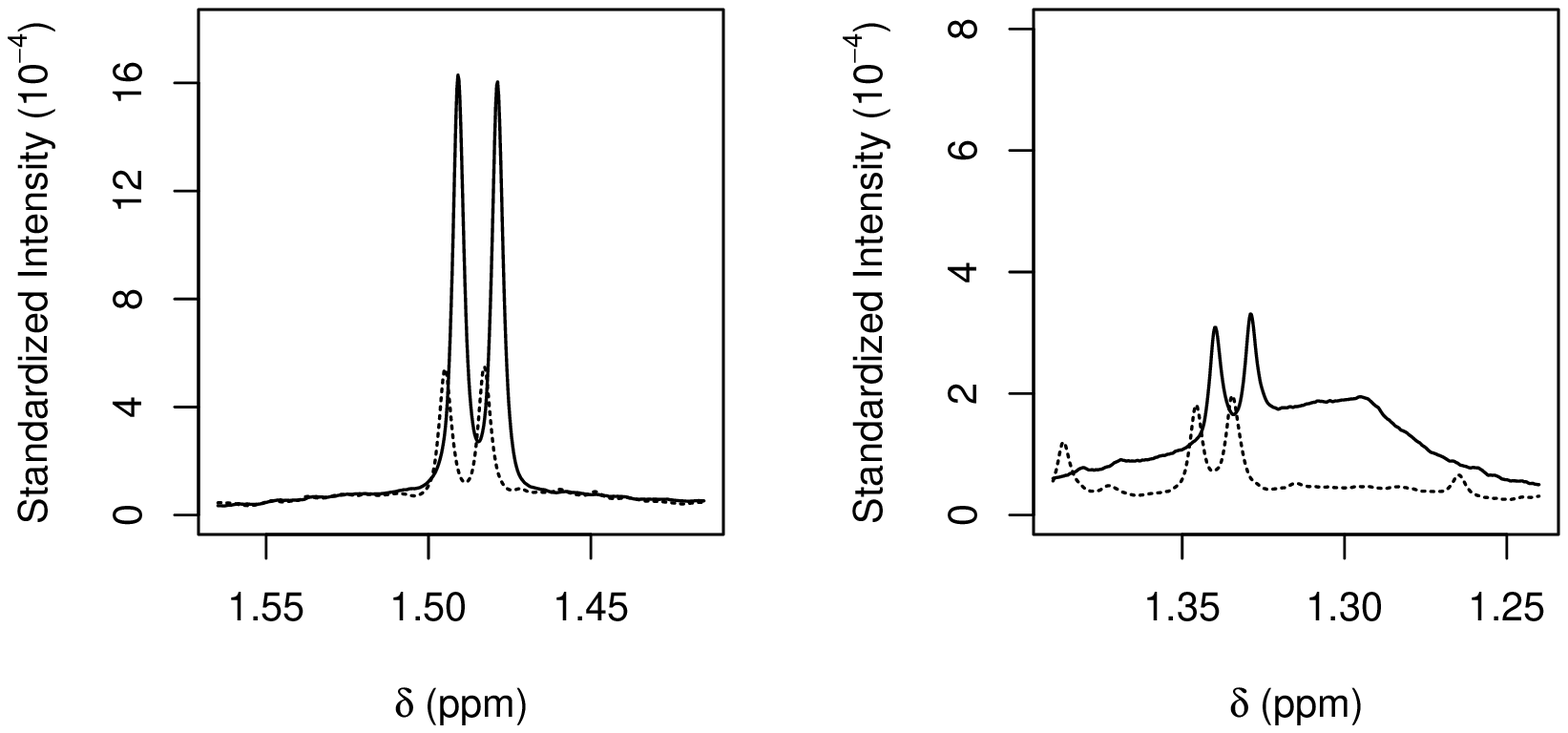}
\end{center}
\caption[Positional noise and peak overlap.]{\label{fig:int:exam_decon} Positional noise between, and peak overlap within, two NMR spectra taken from the yeast experiment; resonances are generated by alanine (\textit{left}) and threonine (\textit{right}).}
\end{figure}

The left panel of Figure \ref{fig:int:exam_decon} illustrates the problem. Excerpts from two spectra corresponding to biological replicates from the same experiment are overlaid, focusing on a doublet type multiplet with two peaks. The difference in peak position between replicates is obvious to the eye. Here, the magnitude of the positional noise is insufficient to confuse assignment by an expert spectroscopist but it will pose problems for standard automated approaches. However, expert deconvolution is rarely practical because it is labor intensive and relies on someone familiar with metabolite resonance patterns. Targeted profiling \cite**{PubMed_16808451} against a standard library of metabolite resonance peaks reduces the importance of expert knowledge but is slow because there is no automated fitting procedure.  Spectral binning (\citeasnoun**{holmes1994automatic}, \citeasnoun**{spraul1994automatic}) approaches divide the spectrum into regions (bins), within which the intensity measurements are averaged, in an attempt to isolate distinct resonance signals. Although this mitigates the effect of peaks fluctuating position within bins, fluctuations across bin boundaries will cause anti-correlated increase/decrease of average intensity in adjacent bins, even if there is no associated change in metabolite concentration. Spectral binning balances parsimony and computational efficiency, retaining quantitative information but in a representation with many fewer, easy to compute, variables. However, the reduced variables are often analyzed without explicit quantification of individual metabolites using pattern recognition methods such as principal components analysis and partial least squares \cite{RIS_0}.

The additional complication of positional noise combined with peak overlap is illustrated in the right panel of Figure \ref{fig:int:exam_decon}. The well defined resonance peaks overlap with broad signals attributable to a combination of closely overlapping low metabolite signals and/or macromolecular signatures. This introduces the problem of estimating the proportion of the signal associated with the sharp resonances and the proportion due to the broad component. The problem becomes even more complex when the target peaks also overlap with other sharp metabolite signals which additionally fluctuate between different spectra.

Currently, there is no statistical methodology that can simultaneously address the problems of identification, deconvolution and quantification when there is positional noise and peak overlap. We believe a method based on explicit quantification from deconvolution of metabolite signatures should have significant advantages: spectral convolution models are parsimonious because they correspond to the physical process generating the data; the variables inferred are interpretable because they represent concentrations of identified metabolites; these concentrations are of direct scientific interest because they depend on the underlying biology.

\subsection{Contributions of this paper}

To tackle the problem of quantifying metabolites in complex biofluids such as cell supernatants or urine, we present a Bayesian model for \textsuperscript{1}H  NMR spectra and a Markov chain Monte Carlo (MCMC) algorithm to automate peak assignment and spectral deconvolution. Bayesian models for NMR data have been described before, notably in \citeasnoun{Bretthorst90bayesiananalysisI}, in \citeasnoun{Bretthorst90bayesiananalysisII}, in many subsequent papers by the same author, in \citeasnoun{dou1996bayesian} and in \citeasnoun{PubMed_17827043}. Our modeling exploits extensive prior information on the resonance signatures of the metabolites, including the expected horizontal displacements and relative vertical scalings of the peaks. This novel approach allows us to deconvolve peaks and assign them to specific metabolites in a unified analysis, which eliminates the need for a manual assignment step. The prior information comes from the physical theory of NMR and from experimental information.  Experimental resonance data on human metabolites are extensive and are publicly available, for example from the online database of the Human Metabolome Project, the HMDB \cite**{PubMed_18953024}.

Almost all biofluid and tissue NMR spectra contain peaks for which there is no prior information in the presently incomplete public metabolite databases. Despite this, the component of a spectrum that cannot be assigned to known compounds, may contain metabolomic information that is scientifically useful, e.g. for classification of spectra. We therefore propose a two component joint model for a spectrum. We model the metabolites whose peaks we wish to assign explicitly parametrically, using information from the online databases, while we model the unassigned spectrum semi-parametrically, using wavelets. We choose wavelets because they model signal continuously but locally. They can account for the local correlation of a spectrum caused by the continuity of the underlying physical processes without imposing unrealistic global modeling constraints.

The wavelet component of our two component likelihood is extremely flexible so that, without restriction, it tends to absorb signal that should be modeled by the parametric component, thus inducing a lack of identifiability. We address this by penalizing the wavelet coefficients using heavy-tailed scale-mixture priors. These priors shrink wavelet coefficients wherever the spectral signal can be explained by the parametric component of the model. We also impose a truncation condition on the wavelets, which reflects prior knowledge that frequency-domain NMR spectra lie almost completely in the upper-half of the $(x,y)$ plane.

To overcome the strong posterior correlation between parameters corresponding to the two model components we introduce purposely designed Metropolis-Hastings block proposals which update the parameters of the two components jointly.

\section{Modelling}\label{sec:met}
\subsection{NMR Spectra}\label{sec:met:NMRSpectra}

Previous authors (\citeasnoun**{Bretthorst90bayesiananalysisI},
\citeasnoun{dou1996bayesian}, \citeasnoun{PubMed_17827043}) developed Bayesian models for NMR data in the time domain, in which resonance signals appear as exponentially decaying sinusoids. However, we prefer to model conventionally preprocessed (by apodization, phase and baseline correction) data in the more interpretable frequency domain, in which resonance signals appear as peaks (e.g. Figure \ref{fig:int:exam_spec}). Our model exploits the positivity of the frequency-spectrum, a condition which cannot be expressed parsimoniously in the time domain. Under an iid Gaussian model for errors, the two representations contain the same information since they are related by an orthogonal transformation (the discrete Fourier transform).

A frequency domain NMR dataset is a pair $(\mathbf{x},\mathbf{y})$, where $\mathbf{x}$ is a length $n$ vector of points on the chemical shift axis, usually regularly spaced and $\mathbf{y}$ is a vector of corresponding resonance intensity measurements. $n$ is typically of the order $10^3$ to $10^4$ depending on the resolution of the spectrum, and the size of the region under consideration.  The intensity measurements are noisy, so that, although they measure inherently positive quantities, some components of $\mathbf{y}$ are likely to fall below the $\delta$-axis.  $\mathbf{y}$ is usually standardized in some way, for example so that $\sum_i^ny_i=1$.

We model $\mathbf{y}|\mathbf{x}$ assuming the $y_i|\mathbf{x}$ are independent normal random variables and,
\begin{equation}
\mathbb{E}\left(y_i|\mathbf{x}\right)=\phi(x_i)+\xi(x_i)
\end{equation}
where the $\phi$ component of the model represents signal from metabolites with peaks we wish to assign explicitly and which have been previously characterized and cataloged in databases. (The metabolites chosen will vary from analysis to analysis according to the prior belief about the content of the biological mixture and the scientific question.) The $\xi$ component of the model represents signal generated by peaks we do not wish to assign (this may include signal from uncataloged resonances of molecules which are partially characterized, with the characterized resonances modeled in the $\phi$ component). We construct $\phi$ parametrically (as a continuous function of continuous chemical shift $\delta$) using the physical theory of NMR, and we model $\xi$ semi-parametrically using wavelets.

\subsection{Modeling $\phi$, the cataloged metabolite signal} \label{sec:met:modphi}

According to physics, the resonance signatures of distinct compounds are independent, accumulate with an intensity proportional to molecular abundance and aggregate in a spectrum by convolution. Consequently, we can write $\phi$ as a linear combination, where each term in the sum corresponds to the signature of one of $M$ different metabolites,

\begin{equation}
\phi(\delta)=\sum_{m=1}^Mt_m(\delta)\beta_m.
\end{equation}
(The value of $M$ will depend on the scientific problem, it is likely to be of order $10^0$ to $10^2$). For each $m$, $t_m$ is a continuous template function which specifies the NMR signature of metabolite $m$. The corresponding coefficient $\beta_m$ is proportional to the molecular abundance of $m$ (i.e. the concentration of $m$) in the biological sample. Physical theory restricts the model space for each template function to a parametric mixture of horizontally translated and vertically scaled Lorentzian peaks \cite{NMRHoreP}.

\begin{figure}
\begin{center}
\includegraphics[width=\textwidth]{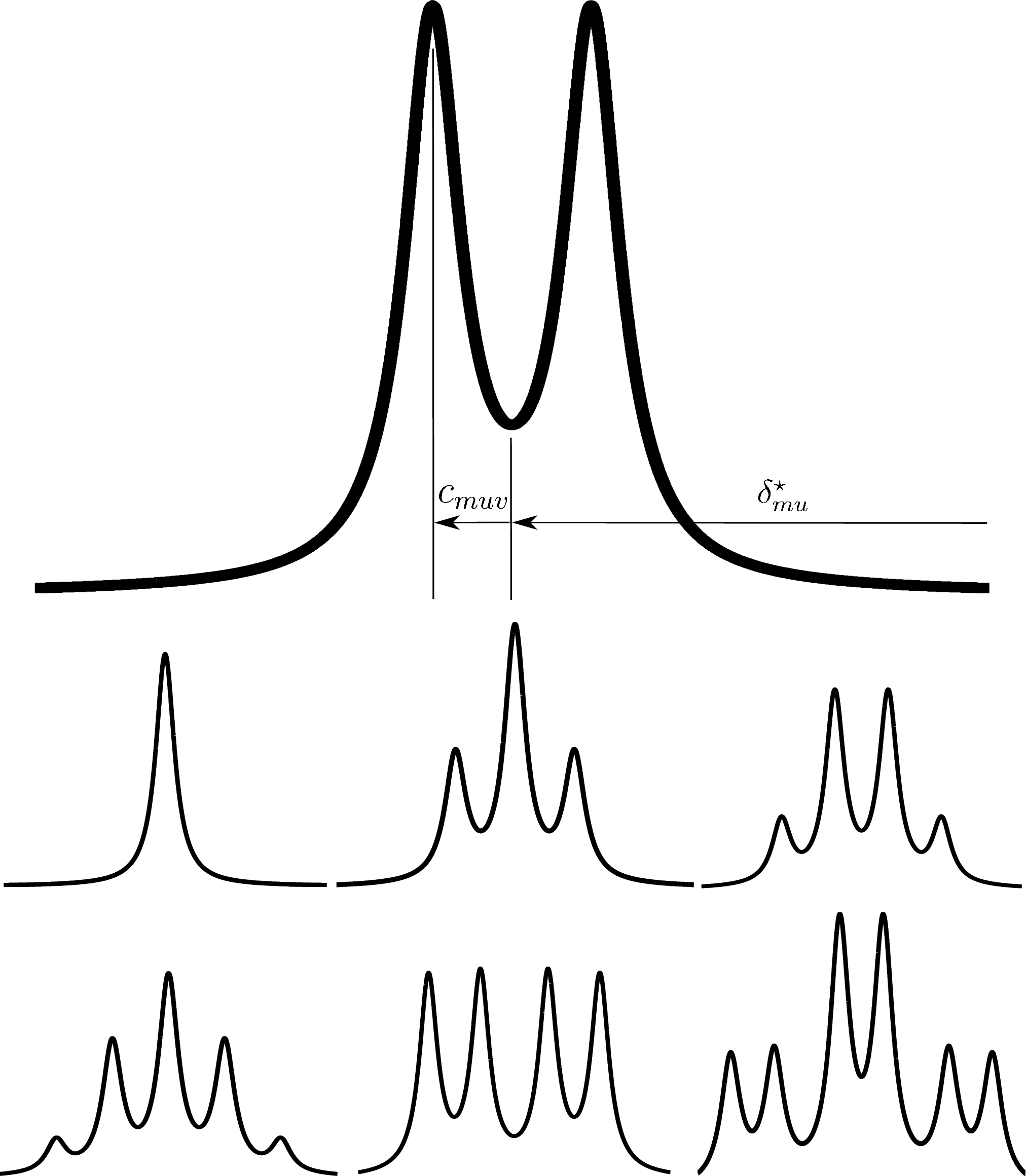}\caption[Some common multiplet types]{\label{fig:mult} The peak configurations of some common types of multiplet. \textit{Top row}: a doublet, with chemical shift $\delta^\star_{mu}$ and peak
offset $c_{muv}$. \textit{Middle row}: (from left to right) a singlet, a triplet, a quadruplet.
\textit{Bottom row}: (from left to right) a quintuplet, a doublet of doublets and a triplet of doublets.}
\end{center}
\end{figure}

We remarked previously that many metabolite signatures contain clusters of Lorentzian peaks called multiplets. Isolated peaks are often classed as \textit{singlet} multiplets and, if we adopt this convention, we can express  each signature template $t_m$ completely as a linear combination of a set of multiplet curves $g_{mu}$,

\begin{equation}\label{eqn:template}
t_m(\delta)=\sum_{u} z_{mu} g_{mu}(\delta-\delta^\star_{mu}),
\end{equation}
where $u$ is an index running over all the multiplets belonging to metabolite $m$. We assume $\int_0^\infty g_{mu}(\delta)d\delta = \int_{-\infty}^0g_{mu}(\delta)d\delta$ for all $m,u$, so that the parameter $\delta^\star_{mu}$ specifies the position on the chemical shift axis of the center of mass of the $u$th multiplet of the $m$th metabolite (see the large multiplet in Figure \ref{fig:mult}). We call $\delta^\star_{mu}$ the chemical shift parameter of the multiplet. Each of the coefficients $z_{mu}$ is a positive quantity, usually equal to the number of protons in a molecule of $m$ that contribute resonance signal to the multiplet $u$. ($z_{mu}$ may sometimes be non-integral, due for example to relaxation effects \cite{NMRHoreP}.  In these cases $z_{mu}$ must be interpreted as an `effective'  proton contribution).  $\int_{-\infty}^\infty g_{mu}(\delta)d\delta$ is constant over $m$ and $u$ so the area under each $t_m$ is proportional to $\sum_u{z}_{mu}$, the number of protons resonating in a molecule of $m$.

With a few exceptions,  multiplets can be classified into one of a number of common types (Figure \ref{fig:mult}) which determine the configuration of the peaks (a doublet, a triplet, a doublet of doublets, etc). This classification together with a small number of continuous quantities called $J$-coupling constants, which determine the (horizontal) distances between the peaks, completely parameterize a multiplet curve.

To be precise, a multiplet curve $g_{mu}$ is the weighted average of $V_{mu}$ translated Lorentzian curves (see eqn. (\ref{eqn:lor})), 
\begin{equation}\label{eqn:multi} 
g_{mu}(\delta)=\sum_{v=1}^{V_{mu}} w_{muv}l_\gamma(\delta-c_{muv}), 
\end{equation}
where the weights $w_{muv}$ (which sum to one over $v$) determine the relative heights of the peaks of the multiplet and the translations $c_{muv}$ determine the horizontal offsets of the peaks from the center of mass of the multiplet (see Figure \ref{fig:mult}). Multiplets are (usually) symmetric so that $\{-c_{muv}:v=1,...,V_{mu}\}=\{c_{muv}:v=1,...,V_{mu}\}$ and $w_{muv'}=w_{muv}$ when $c_{muv'}$=$-c_{muv}$. 

\subsection{Modeling $\xi$, the uncatalogued metabolite signal}

We model $(\xi(x_1), ..., \xi(x_n))^T$ as a linear combination of wavelet basis functions and use $\boldsymbol{\theta}$ to denote the vector of wavelet coefficients. We chose to use Daubechies's least asymmetric wavelets with 6 vanishing moments (\hbox{symlet-6}) as a wavelet basis because these wavelets have a similar shape to Lorentzian peaks. Symlets have been used previously to select features from NMR spectra \cite**{Kim2008161} and sensitivity analysis comparing other potential wavelet bases showed little difference in spectral reconstructions.

\subsection{The Likelihood}\label{sec:like}
We now bring together the models for $\phi$ and $\xi$ to make a formal specification of the probability model for the data. It is easier to do this in the wavelet domain, because the dimension of the wavelet space $p$ often needs to be greater than the dimension of the data space $n$, to deal with distortion at the spectral borders (see \citeasnoun{strang1996wavelets} and Section 1 of the supplementary material.) Let $\mathcal{W}$ be the wavelet transform corresponding to the symlet-$6$ wavelet basis. The likelihood, is defined by
\begin{equation}\label{eqn:errormod}
\mathcal{W}\mathbf{y}=\mathcal{W}\mathbf{T}\boldsymbol{\beta}+\mathbf{I}_p\boldsymbol{\theta}+\boldsymbol{\epsilon},~~~~\boldsymbol{\epsilon}\sim N(0, \mathbf{I}_p/\lambda),
\end{equation}
where $\mathbf{T}$ is the $n\times M$ matrix with $t_m(x_i)$ as its $(i,m)$th entry, $\mathbf{I}_p$ is the $p\times p$ identity matrix and where $\lambda$ is a scalar precision parameter.

Equation (\ref{eqn:errormod}) is a linear regression of $\mathcal{W}\mathbf{y}$ on the columns of $\left[\mathcal{W}\mathbf{T}~\mathbf{I}_p\right]$, the matrix generated by adjoining $\mathcal{W}\mathbf{T}$ and $\mathbf{I}_p$ columnwise. Since this matrix has more columns than rows, the regression coefficients cannot all be identifiable in the likelihood. We address this in the next section, by specifying a prior which helps to distinguish the parametric and semi-parametric components of the model.

\subsection{Prior specification}
Our aim is to obtain a joint Bayesian posterior distribution over the parameters controlling the shape of the templates $\{t_m: m=1,...,M\}$,  and the regression parameters  $\boldsymbol{\beta}$, $\boldsymbol{\theta}$ and $\lambda$; we now specify priors for these parameters.

\subsubsection{\textit{Prior for the peak-width:}}

Our focus is on spectra generated by biofluids such as cell supernatants or urine, for which peak-widths vary between, but negligibly within spectra; it is therefore reasonable to assume that peaks within a spectrum depend on a single common peak-width parameter $\gamma$. Our prior for $\gamma$ is a log-normal distribution, with $\Median(\gamma)=1\textrm{Hz}/F$, $\Var(\gamma)=\textrm{4.6Hz}^2/F^2$ where $F$ is the operating frequency of the spectrometer in MHz. This prior gives good support to a broad region around $1\textrm{Hz}/F$, typical of the peak-widths generated by modern spectrometers \cite{NMRHoreP}. (With this prior, it is easy to relax the assumption of a common peak-width, since local deviations at the metabolite, multiplet, or peak level can be modelled using Gaussian random effects on $\log(\gamma)$.)

\subsubsection{\textit{Prior for the multiplets:}} 

In section \ref{sec:met:modphi} we described a two-level parameterization of metabolite signature templates, defined by (\ref{eqn:template}) and (\ref{eqn:multi}), as a linear combination of Lorentzian peaks nested in multiplets. This allows us to represent a difference in the uncertainty of peak positions within and between multiplets.  The parameters $c_{muv}$ and $w_{muv}$, which determine the multiplet shapes, vary very little across NMR spectra. We assume they are constant and compute them by applying some simple rules (see \citeasnoun{NMRHoreP}, chap. 3 for the details), from empirical estimates of the $J$-coupling constants which are published in online databases. In contrast, as noted in section \ref{sec:challanges}, the multiplet chemical shift parameters $\delta^\star_{mu}$ do fluctuate slightly between spectra according to experimental conditions. We use an estimate $\hat{\delta^\star}_{mu}$ of each $\delta^\star_{mu}$, taken from online databases, to construct an informative prior which accounts for this uncertainty. The positional noise is local and smaller fluctuations are more probable than larger ones, so we assign each $\delta^\star_{mu}$ a truncated normal prior distribution with mean parameter $\hat{\delta^\star}_{mu}$,  variance parameter $10^{-4}\textrm{ppm}^2$ and  truncation region $[\hat{\delta^\star}_{mu}-0.03\textrm{ppm}, \hat{\delta^\star}_{mu}+0.03\textrm{ppm}]$. It may sometimes be appropriate to specify a multiplet or metabolite specific alternative depending on what is known about the variability of particular multiplet locations across spectra.

\subsubsection{\textit{Prior for the metabolite abundances:}}

Having defined a parametric prior for the metabolite signature templates, we now consider the prior for the vector $\boldsymbol{\beta}$, each component of which represents the resonance intensity of a signature and is proportional to the abundance of that metabolite in the biological sample. The intensities are positive so the prior for each component of $\boldsymbol{\beta}$ should confine its support to $\mathbb{R}^+$.  For conditional conjugacy, we assign a normal prior to each $\beta_m$, truncated below at zero, $\beta_m \sim TN(e_m, 1/s_m,0, \infty)$. This is flexible enough to encode prior information for a wide range of research problems. For the simulations and examples presented in this paper we assume low prior information and choose $e_m=0$ and $s_m=10^{-3}$ for all $m$.

\subsubsection{\textit{Prior for the wavelet coefficients and precision parameter:}}

In section \ref{sec:like} we observed that the parametric ($\phi$) and semi-parametric ($\xi$) components of the model are not identifiable in the likelihood. In order to resolve the model components we penalize the semi-parametric component by assigning the wavelet coefficients a prior distribution with heavy tails and a concentration of probability mass near zero. In addition, to reflect prior-knowledge that NMR spectra are mostly restricted to the half-plane above the chemical shift axis, our prior penalizes models in which $\mathcal{W}^{-1} \boldsymbol{\theta}$ has components below a small negative threshold.  To be precise, in order to specify a joint prior for $(\boldsymbol{\theta}, \lambda)$ we introduce a vector of hyper-parameters $\boldsymbol{\psi}$, each component of which corresponds to a wavelet and a vector of hyper-parameters $\boldsymbol{\tau}$, each component of  which corresponds to a spectral data point.   The joint prior for $(\boldsymbol{\theta}, \boldsymbol{\psi}, \boldsymbol{\tau}, \lambda)$ has pdf proportional to

\begin{equation}\label{met:eqn:bigprior}
\lambda^{a+\frac{p+n}{2}-1}\left[\prod_{jk}\psi_{jk}^{c_{j}-{1/2}}\exp\left({-\frac{\psi_{jk}d_j}{2}}\right)\right]\exp\left({-\frac{\lambda}{2}\left(b+\sum_{jk}\psi_{jk}\theta_{jk}^2+r\left(\boldsymbol{\tau}-h\mathbf{1}_n\right)^2\right)}\right)\mathbb{1}_{\{\mathcal{W}^{-1}{\boldsymbol{\theta}}\geq \boldsymbol{\tau} \wedge h\mathbf{1}_n\geq \boldsymbol{\tau}\}}
\end{equation}
The index $jk$ here corresponds to the $k$th wavelet in the $j$th wavelet-scaling level. The following lemma is proved in the supplementary material
\begin{lemma}
Normalization of (\ref{met:eqn:bigprior}) defines a joint prior for $(\boldsymbol{\theta}, \boldsymbol{\psi},\boldsymbol{\tau}, \lambda)$. This prior is proper.
\end{lemma}

The joint prior specified by (\ref{met:eqn:bigprior}) was motivated by consideration of a scale mixture of multivariate normals with smoothed truncation limits:

\begin{equation}\label{met:eqn:thetacond}
P(\boldsymbol{\theta}|\boldsymbol{\psi},\boldsymbol{\tau},\lambda) =
\frac{\lambda^{p/2}\prod_{jk}\psi_{jk}^{1/2}}{C_{\boldsymbol{\psi}\boldsymbol{\tau}\lambda}}
\exp\left(-\frac{1}{2}\sum_{jk}\lambda\psi_{jk}\theta_{jk}^2\right)\mathbb{1}_{\{\mathcal{W}^{-1}{\boldsymbol{\theta}\geq \boldsymbol{\tau}\}}},
\end{equation}
\begin{equation}
\psi_{jk}\sim\Gammad(c_j,d_j/2),
\end{equation}
\begin{equation}\label{met:eqn:priortau}
\tau_i\sim TN(h,1/(\lambda r),-\infty,h),
\end{equation}
\begin{equation}
\lambda\sim\Gammad(a,b/2),
\end{equation}
where $C_{\boldsymbol{\psi}\boldsymbol{\tau}\lambda}$ is a normalizing constant. The index $i$ here corresponds to the $i$th spectral data point. In this specification, $\boldsymbol{\psi}$ allows the prior precision associated with each wavelet to deviate from the global precision $\lambda$. The gamma hyperprior on each component of $\boldsymbol{\psi}$ induces local shrinkage in the marginal prior for $\boldsymbol{\theta}$, which encourages posterior sparsity in the wavelet coefficients. $\boldsymbol{\tau}$ is a vector of $n$ truncation limits, which bounds $\mathcal{W}^{-1}{\boldsymbol{\theta}}$ below. The decaying hyperpriors on the components of $\boldsymbol{\tau}$ smooth these limits and penalize $\boldsymbol{\theta}$ more heavily as more of the semi-parametric component ($\xi$ component) of the model lies below the line $y=h$, where $h$ is a small negative number, chosen close to zero on the spectral intensity scale.

The joint distribution specified by (\ref{met:eqn:thetacond})-(\ref{met:eqn:priortau}) is a reasonable representation of prior belief about $\boldsymbol{\theta}$ and $\lambda$; it places a constraint on the \textit{conditional} distribution of $\boldsymbol{\theta}$ given $\boldsymbol{\psi}$, $\boldsymbol{\tau}$ and $\lambda$.  However, because the normalizing constant $C_{\lambda\boldsymbol{\psi}\boldsymbol{\tau}}$ of (\ref{met:eqn:thetacond}) has no closed form, it is hard to devise a computationally efficient scheme to sample from the resulting `doubly intractable' posterior \cite**{DBLP:conf/uai/MurrayGM06}. The prior defined by (\ref{met:eqn:bigprior})  places the constraint on the \textit{joint} distribution of the parameters, rather than the conditional distribution and it is easy to sample from the full conditionals of $\boldsymbol{\theta}$, $\boldsymbol{\psi}$, $\boldsymbol{\tau}$ and $\lambda$, if we use this prior. We contend that this is an equally valid specification and show in Section 2 of the supplementary material that it behaves similarly to the prior defined by (\ref{met:eqn:thetacond})-(\ref{met:eqn:priortau}).

\begin{figure}
\begin{minipage}[t]{0.5\textwidth}
\begin{center}
\includegraphics[width=0.99\textwidth,clip]{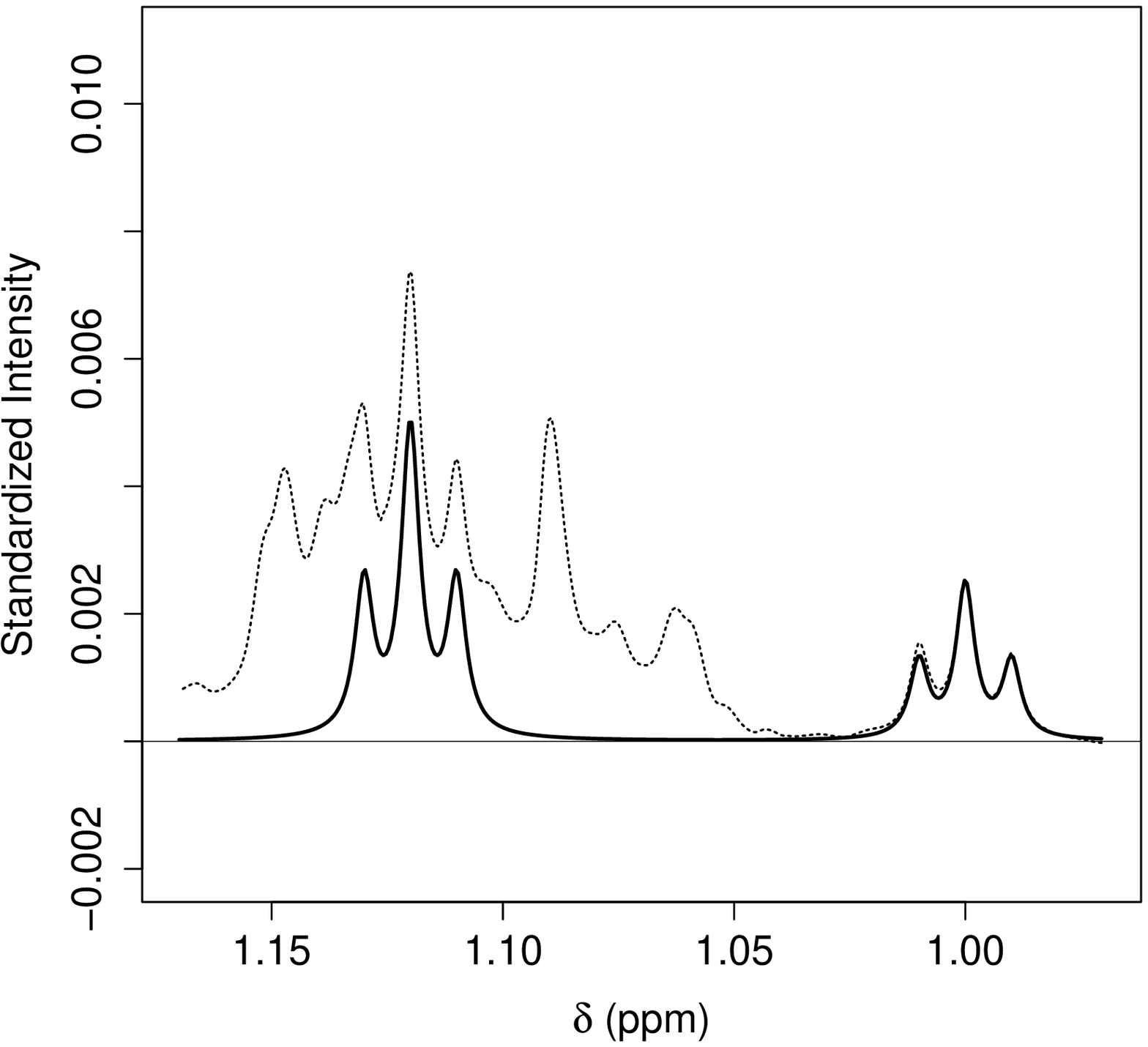}
\end{center}
\begin{center}
\includegraphics[width=0.99\textwidth,clip]{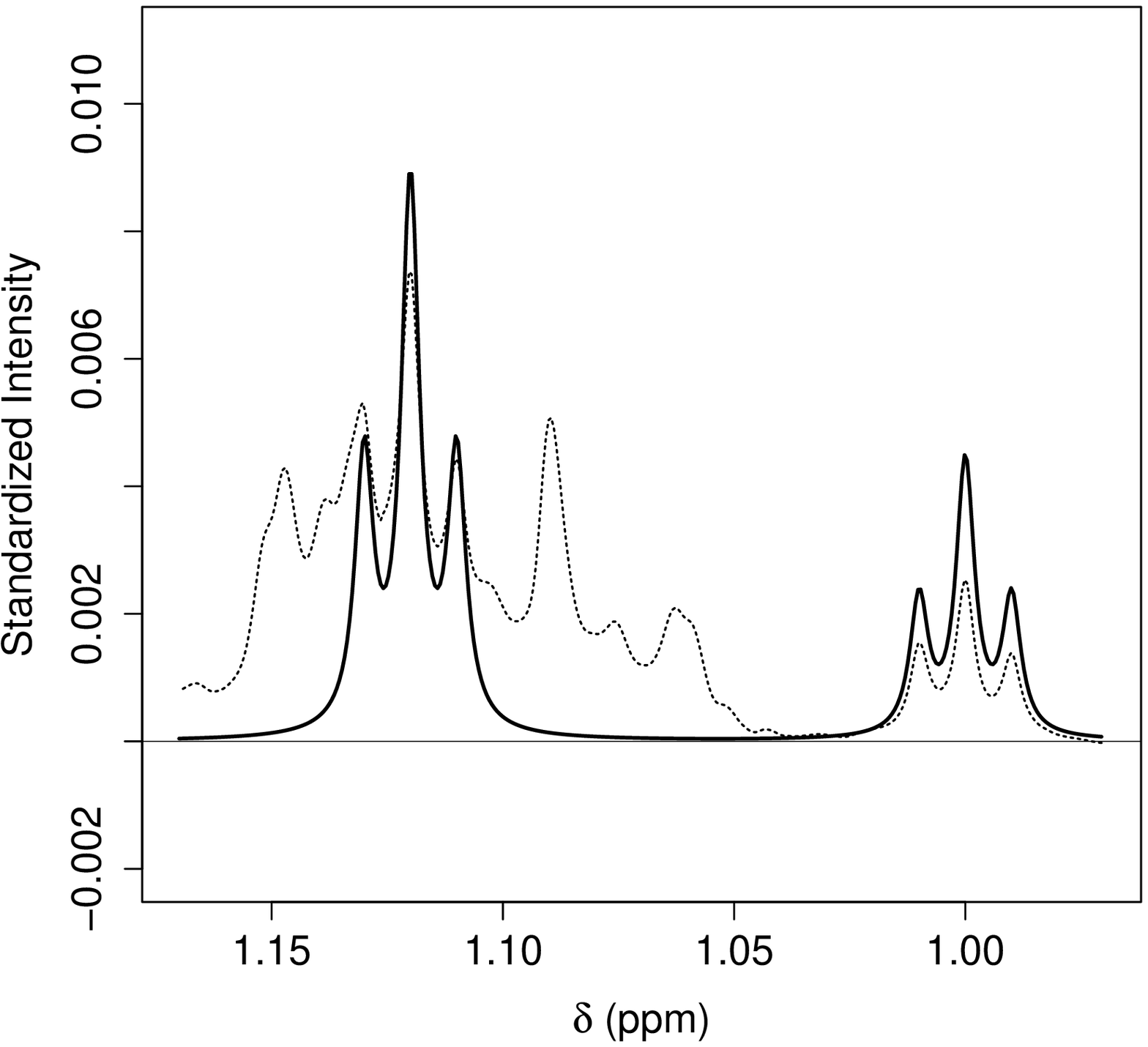}
\end{center}
\end{minipage}
\hfill
\begin{minipage}[t]{0.5\textwidth}
\begin{center}
\includegraphics[width=0.99\textwidth,clip]{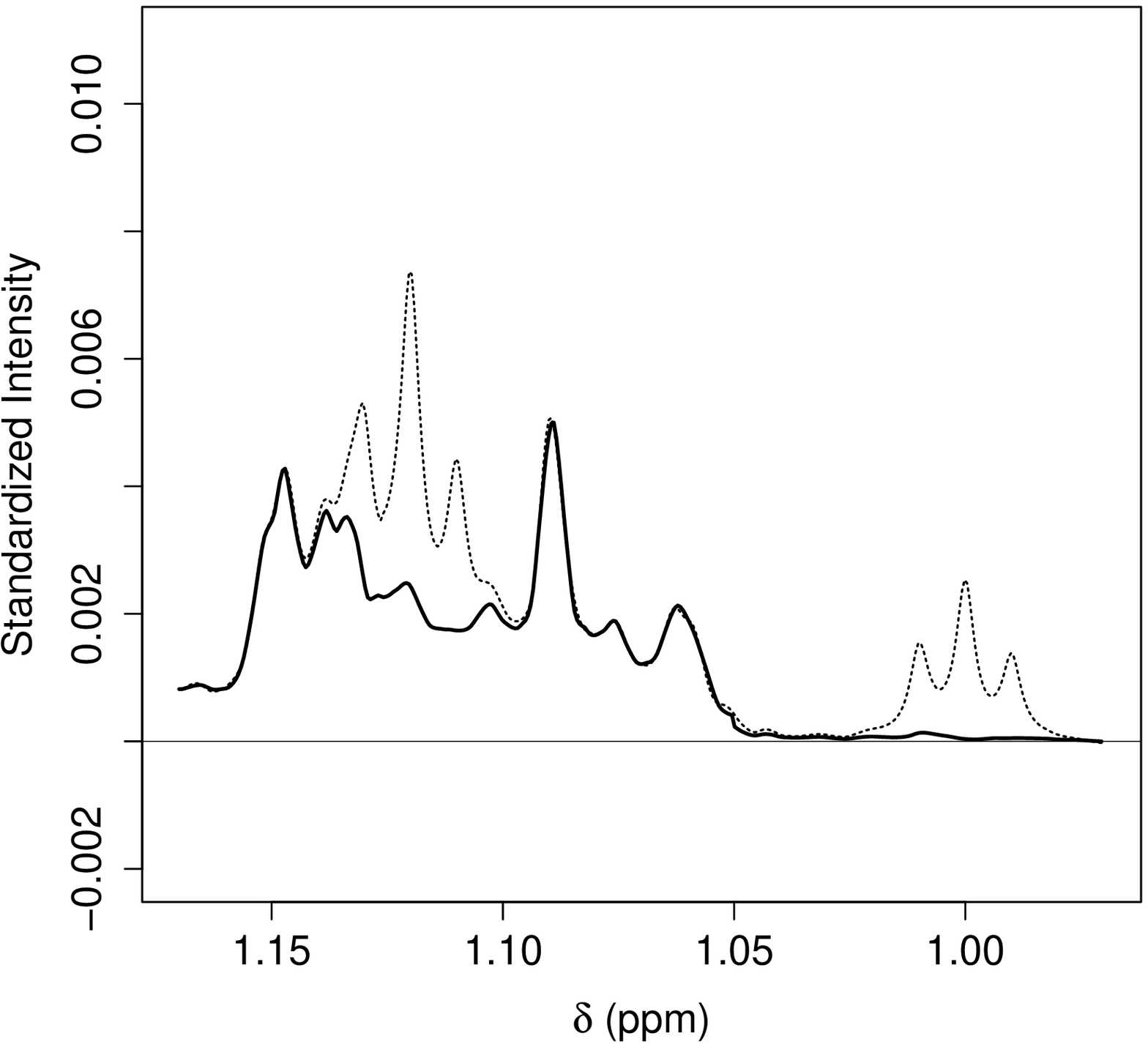}
\end{center}
\begin{center}
\includegraphics[width=0.99\textwidth,clip]{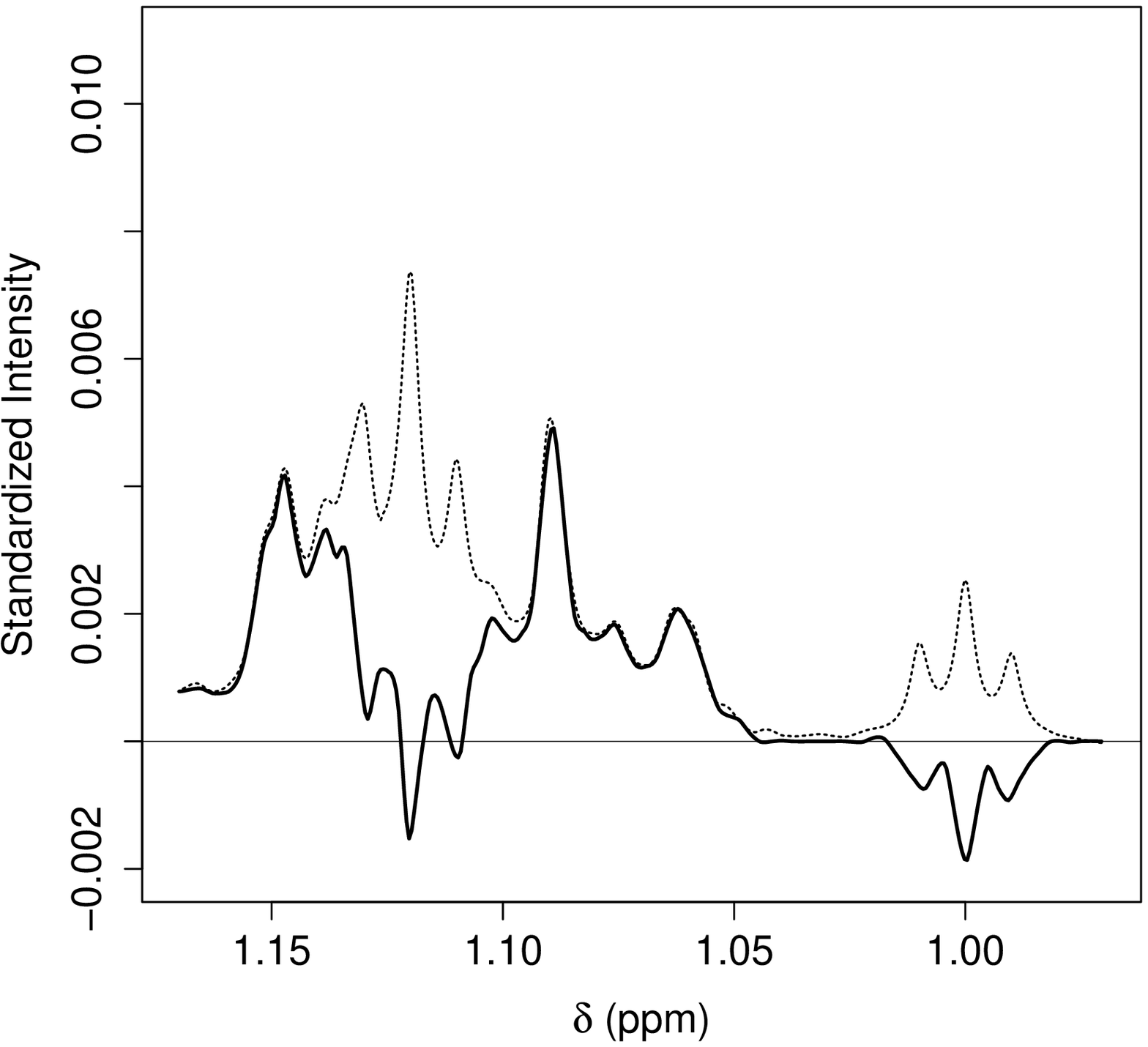}
\end{center}
\end{minipage}
\caption[The effect of a prior penalizing the $\xi$ component in the lower half plane]{\label{fig:trunc}The effect of a prior penalizing the $\xi$ component of the likelihood in the lower half plane (\textit{top}) compared to one without this penalization (\textit{bottom}). The dashed lines show the spectral data.  Deconvolution of a parametric metabolite signature template (heavy lines, \textit{left}) can be more accurate when the wavelet component (heavy lines, \textit{right}) is penalized below the $\delta$-axis.}
\end{figure}

Figure \ref{fig:trunc} demonstrates the effect of penalizing the semi-parametric component of the model when it lies below the chemical shift axis. First, without penalizing $\xi$ in the lower half plane, if the components of $\boldsymbol{\theta}$ are given untruncated {Student's-$t$} priors, the posterior for the vector of quantification parameters $\boldsymbol{\beta}$ focuses asymptomatically on a region close to the ordinary least squares estimate of the parameters, $(\mathbf{T}^T\mathbf{T})^{-1}\mathbf{T}^T\mathbf{y}$, while the wavelet component absorbs most of the residual spectrum. When a metabolite has a multiplet embedded in a region of unassigned spectral resonance the least squares estimate overestimates the corresponding quantification parameter. The signature templates absorb spectral signal even when they do not match the shape of the spectral data and this leads to strong posterior support for negative values for some components of $\mathcal{W}^{-1} \boldsymbol{\theta}$. The prior defined by (\ref{met:eqn:bigprior}) however, can correct the concentration estimate providing at least one of the multiplets of the metabolite is deconvolved cleanly.

\subsubsection{\textit{Scale-invariant inference and hyper parameter settings:}}

Note that in the limit $a\rightarrow 0, b\rightarrow 0, \forall m,  s_m \rightarrow 0$, the (improper)  prior is invariant under the scaling reparameterisation $(\boldsymbol{\beta},\boldsymbol{\theta},\lambda)\mapsto(S\boldsymbol{\beta},S\boldsymbol{\theta}, \lambda/S^2)$ for every constant $S>0$ so that, in this limit, inference is unaffected by the scale of measurement of the data. Smaller $a$ and $b$ correspond to increased uncertainty in the value of $\lambda$. For simulations and examples described in this paper we take $a=10^{-9}$ and $b=10^{-6}$.

The values of the $c_j$ and $d_j$ control the degree of shrinkage penalization imposed on the wavelet coefficients.  Experience shows $c_j=0.05$ and  $d_j=10^{-8}$ provides adequate penalisation. More stringent penalization is possible but our MCMC algorithm tends to mix less well as the penalty gets stronger, because our block updates are less good at targeting the true posterior distribution of $\theta$ (see below). $h$ controls the threshold below which the wavelets are penalised in the lower half plane while $r$ controls the strength of that penalisation. We choose $h$ a little below zero to be sure that true signals below $y=0$, due for example to baseline wiggle can be adequately modelled by wavelets. We set $h=-0.002$ and $r=10^5$.

\section{Markov chain Monte Carlo Algorithm}\label{sec:MCMC}

We implemented a Markov chain Monte Carlo algorithm, to sample from the joint posterior distribution of the model parameters. There are three types of MCMC update.
\begin{itemize}
\item Firstly, there are Gibbs samplers for the components of $\boldsymbol{\beta}$ (truncated normal), the components of $\boldsymbol{\theta}$ (truncated normal),  the components of $\boldsymbol{\psi}$ (gamma), the components of $\boldsymbol{\tau}$ (truncated normal) and $\lambda$ (gamma). The specific distributions for these updates are given in the supplementary material.
\item Secondly, there is a Metropolis-Hastings update for each of  the parameters (except the components of $\boldsymbol{\beta}$) controlling the parametric component of the model.

Specifically, in order to update the multiplet chemical shift parameter $\delta^\star_{mu}$ we propose ${\delta^\star_{mu}}'$ from
\begin{equation}
TN(\delta^\star_{mu},V_{\delta^\star_{mu}}^2  \hat{\delta^\star}_{mu}-0.03\textrm{ppm}, \hat{\delta^\star}_{mu}+0.03\textrm{ppm}),
\end{equation} 
a Gaussian proposal, centred on the current parameter value and truncated at the boundaries of the prior distributions.  We adapt the proposal variance $V_{\delta^\star_{mu}}^2$ using the Adaptive Metropolis-within-Gibbs algorithm of \citeasnoun**{roberts2009examples}. $V_{\delta^\star_{mu}}$ is tuned to target an acceptance rate of 0.44 by increments and decrements which decay in magnitude asymptotically like the inverse of the square root of the iteration number.

In order to update the peak-width parameter we make a Gaussian proposal for $\log \gamma$, centred on the current parameter value. Again, we adapt the proposal variance following \citeasnoun**{roberts2009examples}.

The likelihood constrains $\mathcal{W}\mathbf{y}-\mathcal{W}\mathbf{T}\boldsymbol{\beta}-\boldsymbol{\theta}$, inducing strong posterior correlation between the parametric and semi-parametric components of the model. Consequently, updates of this type, which just propose changes to the parametric component,  can only make local moves in the state space of the Markov chain.

\item Thirdly, there are Metropolis-Hastings block updates in each of which a parameter controlling the parametric component of the model is updated jointly with the vector $\boldsymbol{\theta}$ of wavelet coefficients. The joint proposal breaks the posterior correlation between the parametric and semi-parametric model components, allowing larger jumps in state-space.

Block updates of the $\delta^\star_{mu}$ extend the univariate proposals described previously. First we draw the univariate proposal $\delta'^\star_{mu}$ (although we fix the proposal variance, see supplementary material), then conditional on the value drawn we propose a new value $\boldsymbol{\theta}'$ for the vector of wavelet coefficients $\boldsymbol{\theta}$ and perform a global Metropolis-Hastings accept-reject assessment for the block update.

The conditional proposal for $\boldsymbol{\theta}'$ is a multivariate truncated normal distribution with mean parameter $\mathcal{W}\mathbf{y}-\mathcal{W}\mathbf{T'}\boldsymbol{\beta}$ (where $\mathbf{T}'$ is the template matrix updated to reflect the initial univariate proposal), precision parameter $\lambda\mathbf{I}_p$ and truncation $\mathcal{W}^{-1}{\boldsymbol{\theta}'}\geq \boldsymbol{\tau}.$  We can simulate from this distribution by making the change of basis, $\boldsymbol{\eta}'=\mathcal{W}^{-1}\boldsymbol{\theta}'$; the components of $\boldsymbol{\eta}'$ are then independent univariate truncated normal distributions. This choice of conditional proposal is motivated by the full conditional of $\boldsymbol{\theta}$,

\begin{equation}\label{fullcontheta} \boldsymbol{\theta}|\boldsymbol{\psi},\boldsymbol{\tau}, \lambda, \mathbf{y} \sim TMVN \left(\mathcal{W}\mathbf{y}-\mathcal{W}\mathbf{T}\beta, \left(\mathbf{I}_p+\boldsymbol{\Psi}\right)^{-1}/\lambda, \mathcal{W}^{-1}{\boldsymbol{\theta}}\geq \boldsymbol{\tau}\right)
\end{equation}
which has  a similar distribution but with a reduced precision. Unfortunately there is no easy way to sample from (\ref{fullcontheta}) because the truncation condition $\mathcal{W}^{-1}{\boldsymbol{\theta}}\geq \boldsymbol{\tau}$ induces a complex dependence structure between the components of $\boldsymbol{\theta}$.

The only proposals not yet described are those for the block updates of each component of $\boldsymbol{\beta}$ with $\boldsymbol{\theta}$.  We propose $\beta_m'$ from a Cauchy distribution, truncated below at zero. We center the Cauchy distribution on the point that maximizes the full conditional of $\beta_m$ subject to $\boldsymbol{\theta}=\mathbf{0}$ and to the truncation condition $\mathbf{y}-\mathbf{T}\boldsymbol{\beta}>\boldsymbol{\tau}.$ This is a greedy proposal, in the sense that it attempts to maximise $\beta_m'$ and explain as much of the spectral signal as possible using the template for metabolite $m$, excluding the wavelets ($\boldsymbol{\theta}=\mathbf{0}$). Conditional on the proposed $\beta_m'$ we propose $\boldsymbol{\theta}'$ from a multivariate truncated normal distribution with mean parameter $\mathcal{W}\mathbf{y}-\mathcal{W}\mathbf{T}\boldsymbol{\beta}'$ (where $\boldsymbol{\beta}'$ is $\boldsymbol{\beta}$ with the $m$th component set to $\beta_m'$), precision parameter $\lambda\mathbf{I}_p$ and truncation $\mathcal{W}^{-1}{\boldsymbol{\theta}'}\geq \boldsymbol{\tau}.$

 \end{itemize}

Although, on average, block updates move the chain further than the corresponding single parameter updates, the acceptance rate is lower. Consequently moves of the second and third type make complementary contributions to MCMC mixing.

\subsubsection{\textit{Improving convergence and mixing:}}During the burn in stage of the MCMC we temper the likelihood and penalize the wavelet component of the model stringently to help the chain move into a region of good posterior support.  A parameter $T$ (see section $2$ of supplementary material), which jointly controls the temperature/penalization, gradually cools according to a deterministic schedule, proportional to the complement of a Gaussian cdf.

To improve mixing, we implemented a population (multi-chain) version of our algorithm in which the MCMC operates on a product state-space composed of copies of the space described in Section \ref{sec:met}.  The chain targets a tempered version of the posterior distribution of interest marginally on each subspace, but with each subspace taking a different value of $T$.  The Metropolis-Hastings updates already described operate within each subspace; the acceptance probabilities for these depend only on the state of the relevant subchain and so they can be carried out in parallel on different CPU cores. Additionally, there are two types of update which allow the transfer of information between subspaces. Firstly, we propose that subspaces adjacent in the ordering of $T$ swap parameter values, as in the exchange moves of parallel tempering \cite*{Geyer:1991tj}.  Secondly, for a multiplet location parameter $\delta^\star_{mu}$ in a given subchain, we pick a complementary subchain uniformly at random  and propose a new value for $\delta^\star_{mu}$ from a Cauchy distribution, centred on the value of the parameter in the complementary chain, with scale-parameter $3\textrm{ppm}$. We then propose a conditional update for the $\boldsymbol{\theta}$ of the given subchain in the manner of the block-updates already described.  This proposal is made for every chemical shift parameter of every subchain once for each iteration of the MCMC algorithm. It allows good values of a chemical shift parameter to spread through the population of chains. In principle additional information sharing moves (e.g. Evolutionary Monte Carlo crossover \cite{Liang:2000ws}) could be added, but we find this `copy' move to be sufficient. Ergodicity and the marginal convergence of the subchain with $T=1$ to a stationary distribution equal to the target posterior is assured by theory \cite{Jasra:2007id}.  We combine annealing with parallel chains by running a ladder in which the ratios of the temperatures of the subchains are constant over time, and for which the target subchain (that with $T=1$) cools according to the specified annealing schedule.

\section{Performance}

We simulated $100$ biofluid NMR spectra by convolving the standard library spectra for pure samples of acetic acid, alanine, betaine, creatine, glycine, glycolic acid, guanidoacetic acid, lactic acid, succinic acid, taurine, trimethylamine and trimethylamine oxide all of which generate resonances in the region $0\textrm{ppm}-5\textrm{ppm}$. The concentration of each metabolite was generated as a $U[0,1]$ random variable and the multiplets of each metabolite were perturbed by draws from $U[-0.03\textrm{ppm},0.03\textrm{ppm}]$. We mixed each simulated spectrum with a broad Gaussian pdf the mean of which was drawn uniformly from the range of the spectrum ($0\textrm{ppm}-5\textrm{ppm}$) and standard deviation of which was fixed at $5.8\textrm{ppm}$. This hump is typical of the broad baseline distortions commonly generated by macromolecules in biofluid spectra.

Because each spectrum in the standard library was generated from a different
\textsuperscript{1}H NMR experiment, the width of the peaks in a simulated spectrum can vary slightly between resonances generated by different compounds. This variability can affect inferences of concentration made with our model because of correlation between concentration and peak-width parameters. To deal with this artefact of the simulator, we extended the model by adding random effects to allow for some (small) inter-metabolite variability in the peak-width.

We applied the (single chain version of the) MCMC algorithm for the extended model to each simulated spectrum, running for $5~000$ iterations with a $3 000 $ iteration burn in. \subsection{Multiplet Localization}

\begin{figure}
\includegraphics[width=0.99\textwidth]{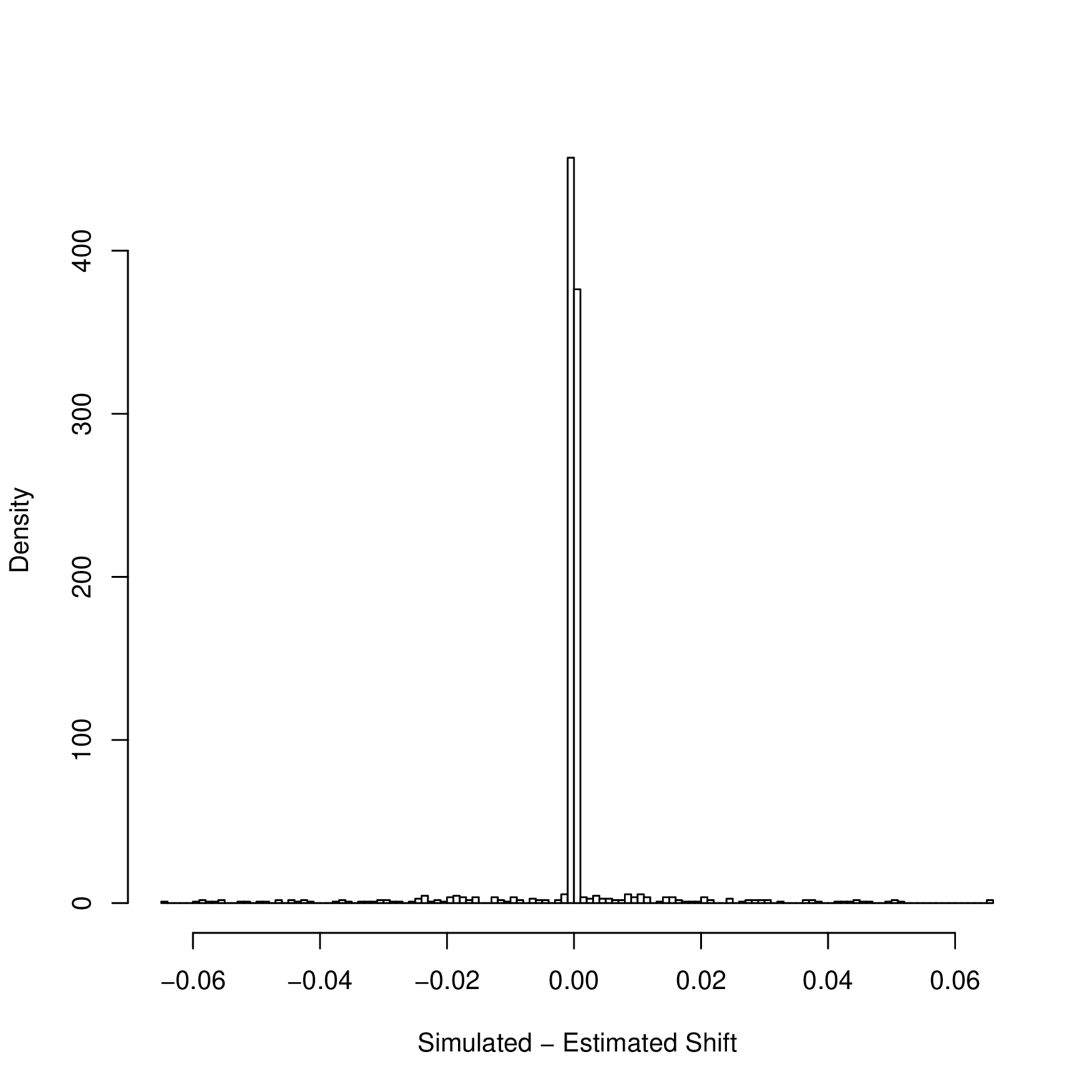}\caption[Bayesian estimated multiplet shifts vs simulated shifts]{\label{fig:int:shifts} Normalized histogram of simulated minus Bayesian posterior mean estimates of multiplet shift.}
\end{figure}

Figure \ref{fig:int:shifts} is a normalised histogram of the difference between the Bayesian posterior mean estimate of each multiplet chemical shift and the true simulated chemical shift, illustrating the quality of the peak localisation. The Bayesian posterior mean estimates $85\%$ of shifts within $0.002\textrm{ppm}$ and $90\%$ within $0.015\textrm{ppm}$.

\subsection{Quantification}

We compared our Bayesian method for quantification to the following algorithm for numerical integration, which is slightly more sophisticated than conventional spectral binning because it includes a peak identification step. To integrate a multiplet we used the known simulated perturbation in chemical shift to identify the precise location of the multiplet. In practice, this information would be unknown to a spectroscopist, at best he or she might be able to identify a multiplet by eye and choose bin boundaries based on the observed location. The information is, of course, unavailable to the Bayesian method. Using the pure compound spectrum we identified the region $[L,R]$ of the chemical shift axis that corresponds to the central $95\%$ mass of the multiplet's parametric template. Finally, we estimated the total area under the multiplet by $\left(\sum_{L<x_i<R} y_i\right)/(0.95N(L-R))$, where $N$ is the number of $x_i$ in $[L,R]$.

\begin{figure}\begin{minipage}[t]{0.5\textwidth}
\includegraphics[width=0.99\textwidth,clip]{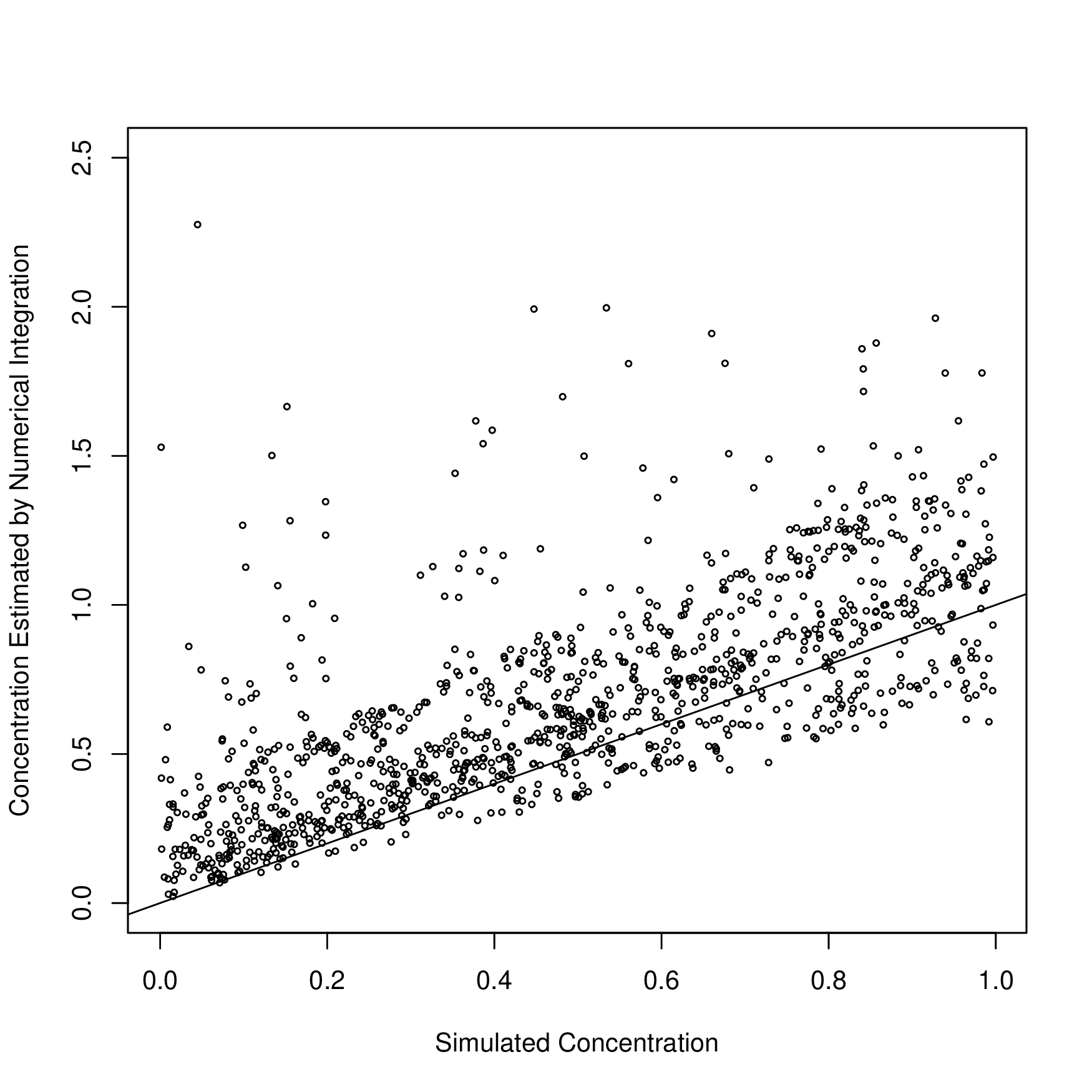}
\end{minipage}
\begin{minipage}[t]{0.5\textwidth}
\includegraphics[width=0.99\textwidth,clip]{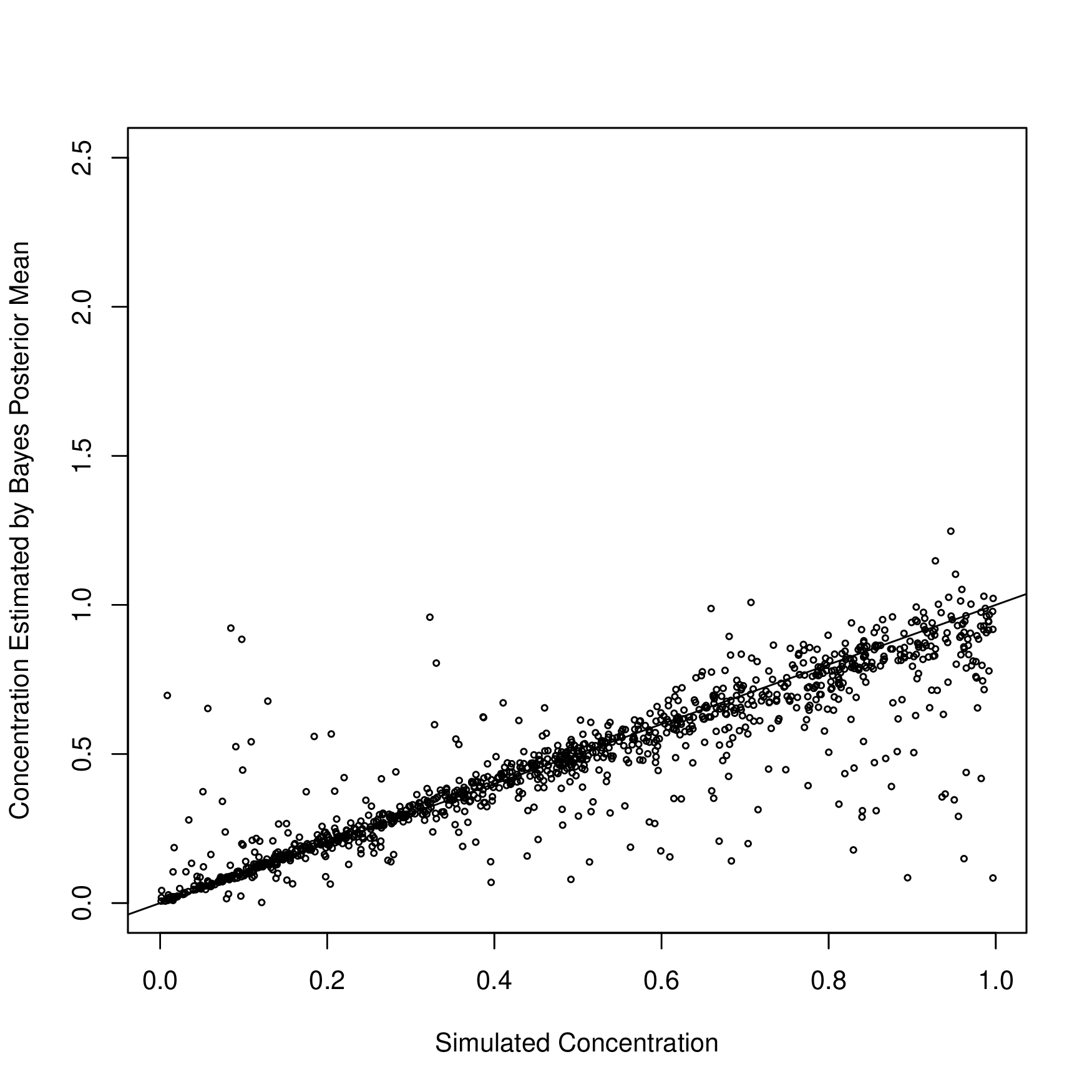}\end{minipage}
\caption[Comparison of concentration estimates from simulated spectra. ]{Concentration estimates from simulated spectra: true vs estimated concentrations for numerical integration (\textit{left}) and for Bayesian posterior mean (\textit{right}). \label{fig:int:simconcs}}
\end{figure}

The plots in Figure \ref{fig:int:simconcs} show the actual vs. estimated concentrations for estimates made by numerical integration and Bayesian posterior mean. The numerical method is systematically biased towards overestimation for two reasons. Firstly, it fails to take account of the Gaussian background hump and secondly it cannot distinguish resonance in the region $[L,R]$ generated by the metabolite of interest from other confounding signals. The Bayesian estimate is unaffected by the first problem because the background hump is modeled with wavelets;  the second problem is dealt with by deconvolution.   The mean quadratic error for the numerical estimator is $0.106$ (equivalent to Pearson correlation 0.70), where the mean is taken over all simulation replicates and metabolites. In comparison, the quadratic error for the Bayesian posterior estimator is $0.017$ (equivalent to Pearson correlation 0.89), over sixfold smaller.

To investigate the performance of the Bayesian method under different strengths of multiplet chemical shift perturbation we ran an additional $150$ replicates of the simulations previously described in $3$ blocks of $50$ with perturbations drawn from $U[-0.01\textrm{ppm},0.01\textrm{ppm}]$, $U[-0.02\textrm{ppm},0.02\textrm{ppm}]$  and $U[-0.04\textrm{ppm},0.04\textrm{ppm}]$ in each block.  We extended the truncation on the chemical shift priors to $\pm 0.045 \textrm{ppm}$ the database estimate. The peak assignments and concentration estimation errors are summarized in Table \ref{perttab}. There is no evidence of a major trend in the quality of peak assignment or concentration estimation with the magnitude of multiplet perturbation.

\begin{table}[Peak assignment and estimation errors.]
\caption{Peak assignment and concentration estimation errors for different magnitudes of simulated chemical shift perturbation.\label{perttab}}
\begin{tabular}{|p{4cm}|p{3.2cm}|p{3.2cm}|p{3.2cm}|}\hline
 Distribution of Simulated Perturbations& Peaks Assigned Within $0.002\textrm{ppm}$  &  Mean Quadratic Error for Bayes Estimate& Mean Quadratic Error for Numerical Integration Estimate\\ \hline
   $U[-0.01\textrm{ppm},0.01\textrm{ppm}]$ &$83\%$&$0.016$& 0.141\\\hline
   $U[-0.02\textrm{ppm},0.02\textrm{ppm}]$ &$84\%$&$0.016$& 0.102\\\hline
   $U[-0.04\textrm{ppm},0.04\textrm{ppm}]$ &$83\%$&$0.017$& 0.135\\\hline
\end{tabular}
\end{table}
\section{Yeast Dataset}
To illustrate the performance of our method on a real dataset, we took three spectra from the experiment investigating the metabolic response of the yeast \textit{Pichia pastoris} to recombinant protein expression \cite**{PubMed_21283710}.  The spectra were generated using biological replicates prepared under the same conditions. Consequently, the metabolic profiles of the samples are extremely similar and the spectra contain essentially the same metabolite concentration information. Nevertheless, the spectra are slightly different, because for example, of experiment level positional noise in the chemical shifts of resonance peaks. By modeling the three spectra jointly we can quantify metabolites using information from all three replicates, while accounting for these experiment level differences.

We used the model described in Section \ref{sec:met} as a basis for a joint model of multiple spectra.  In the new model, the vector of metabolite quantification parameters $\boldsymbol{\beta}$ is held in common across the spectra. All the remaining parameters are copied from the original model, with a replicate set assigned to each spectrum. The MCMC algorithm for the multiple-spectra model is very similar to the procedure described for the original model. The Metropolis-Hastings updates involving components of $\boldsymbol{\beta}$ need to be adjusted, to reflect the dependence on multiple spectra, but are similar to those for the simpler model (see section 3 of the supplementary material). The updates for the remaining parameters continue to be valid within each spectrum because, conditional on $\boldsymbol{\beta}$, the joint posterior factorizes into separate probability models, each corresponding to a different spectrum.

\citeasnoun{Tredwell:2011kr} manually quantified $37$ metabolites from these spectra, with each of the five authors assigning the resonances and estimating concentrations independently. Not all the resonance patterns generated by the $37$ metabolites take the form of the symmetric multiplets described in Section \ref{sec:met}. Multiplet shapes are sometimes distorted by strong interaction effects, which we cannot easily include in our model because they are neither described by a known parametric model nor cataloged in a public database. However, distorted multiplets are still convolutions of Lorentzian peaks, so it is sometimes possible to construct a template-based model by estimating the weights and translations of (\ref{eqn:multi})  from an NMR spectrum of the relevant pure compound generated under similar experimental conditions. We were able to construct a parametric signature template for 26 of the $37$ compounds by combining public database information with parameter estimates from our laboratory library of pure compound spectra. However, we were unable to construct complete signatures, containing a full complement of multiplets, for every metabolite. Although this precludes a complete deconvolution of the spectral signal generated by compounds with unmodeled resonances, and the omitted resonance signals will be absorbed into the wavelet component of the model. Nevertheless, our main aim is to obtain accurate concentration estimates and this is still achievable, providing at least one multiplet from each metabolite deconvolves correctly.

\begin{figure}\begin{center}
\includegraphics[width=0.99\textwidth,clip]{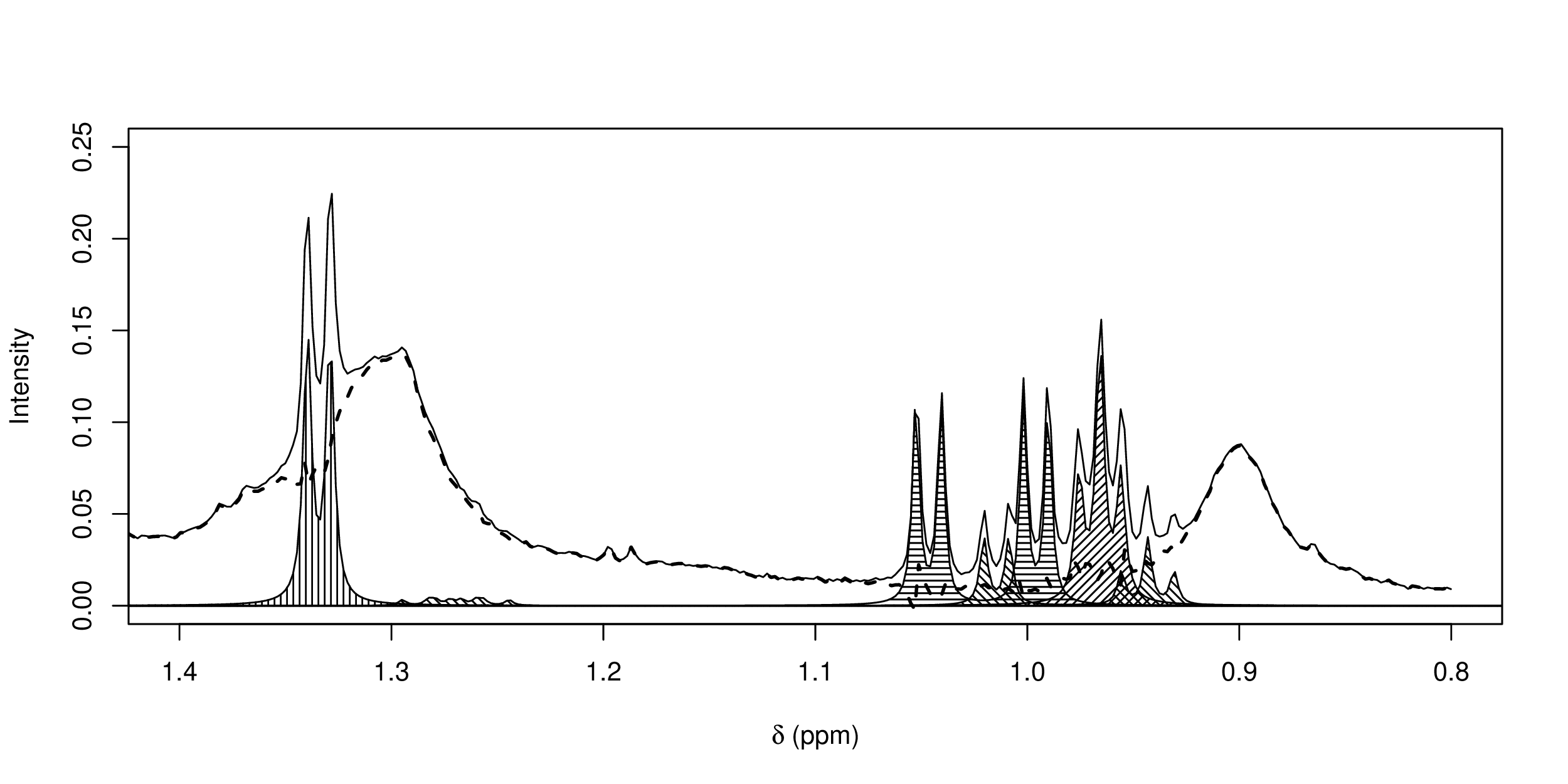}
\end{center}
\begin{center}
\includegraphics[width=0.99\textwidth,clip]{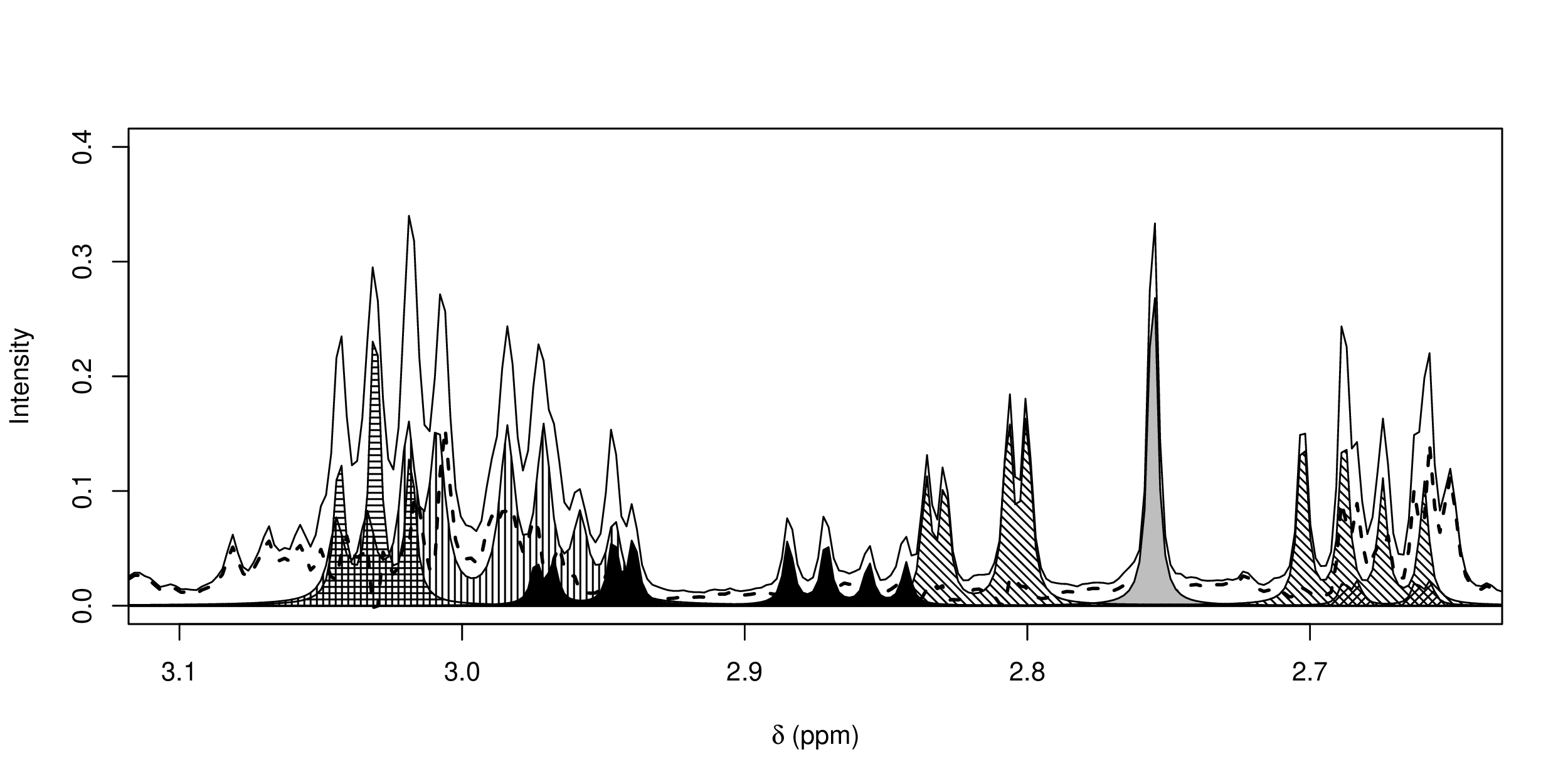}
\end{center}\caption[Deconvolutions of the yeast spectra]{\label{fig:int:deconvreg}Deconvolution of selected regions from one of the yeast spectra. The top panel shows resonances generated by isoleucine (NW-SE hatch), leucine (NE-SW hatch), threonine (vertical hatch) and valine (horizontal hatch). The lower panel shows resonances generated by aspartate (NW-SE hatch), asparagine (black shade), histidinol (vertical hatch), lysine (horizontal hatch), malate (NE-SW hatch) and methionine sulfoxide (grey shade). The deconvolution is conditional on the MAP estimates of the peak-width and chemical shift parameters  and plotted on the same grid as the original spectrum. The original spectral data is shown by the continuous lines and the wavelet component of the model by the dashed lines.}
\end{figure}

We ran the MCMC procedure, with $8$ parallel chains tempered on a ladder, for $20~000$ iterations following a $10~000$ iteration burn in. We made an adjustment to the prior on the chemical shift parameters of the singlet multiplets generated by Histidinol near to $7.25\textrm{ppm}$ and $8.19\textrm{ppm}$ by truncating the prior at $\pm 0.5\textrm{ppm}$ of the HMDB estimate rather than at $\pm 0.03\textrm{ppm}$. This represents prior knowledge that the chemical shifts of those multiplets are more variable than is typical because of sensitivity to chemical properties of the biofluid.  Figure \ref{fig:int:deconvreg} shows the posterior deconvolution of a heavily congested region ($2.6\textrm{ppm}-3.1\textrm{ppm}$) and a region of broad ($0.8\textrm{ppm}-1.4\textrm{ppm}$) from one of the spectra.

\begin{figure}\begin{center}
\includegraphics[width=0.99\textwidth,clip]{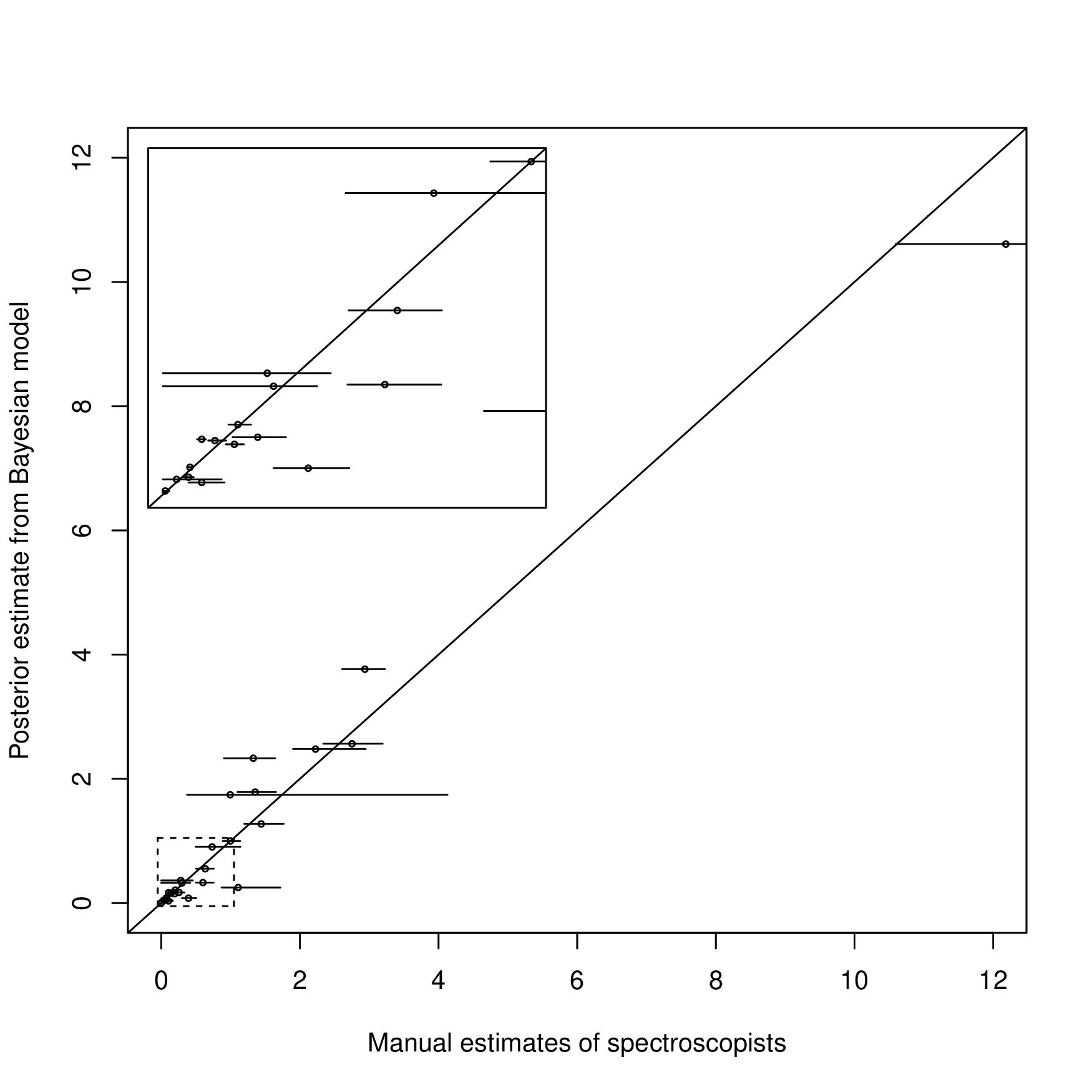}
\end{center}
\caption[Comparison of concentration estimates]{\label{fig:int:conccomp} Comparison of metabolite concentration estimates by posterior mean of the quantification parameters from the Bayesian model ($y$-axis) with manual spectroscopists estimates ($x$-axis). The scales are calibrated so that the concentration of Trehalose is equal to 1.   Each circle represents the mean of the five spectroscopists estimates while the limits of the horizontal bars represent the range. The inset is a magnification of the region bounded by the dashed line.}
\end{figure}

The five spectroscopists deconvolved the spectra with the assistance of the widely used Chenomx spectrographic software, which implements a form of targeted profiling \cite{PubMed_16808451}. This is probably the most precise method currently available for estimating metabolite concentrations from spectra, although its accuracy depends on spectroscopists being able to make correct peak assignments. Figure \ref{fig:int:conccomp} is a plot showing the strong concordance between the concentration estimates of the spectroscopists and concentration estimates made by the Bayes posterior mean. Although $11$ of the $26$ posterior mean estimates lie outside the range of the spectroscopists' estimates, there are only $3$ cases of substantial discordance. The statistical estimates for arginine and histidinol are substantially larger than the spectroscopist's estimates while that for malic acid is substantially lower.  In the cases of arginine and histidinol it is hard to fault the posterior deconvolution by visual inspection. The discrepancies could be due to an error in the templates used by the spectroscopists to profile the resonances or to experimental errors in the database estimates of the parameters used to construct the metabolite signature templates of Bayesian model. In the case of malate the discrepancy appears to have been caused by a multiplet misalignment. In principle this could be resolved by adjusting the prior for that multiplet's chemical shift parameter in order to force the correct alignment.

\section{Discussion}

Presently, automatic methods for analyzing biofluid NMR data rely on non-parametric pattern recognition techniques or are based on approximate numerical integration algorithms, such as spectral binning. These methods ignore a large amount prior information about the physical process generating the spectral data.

Prior information about a data generating process can easily be incorporated into a Bayesian analysis through specification of a likelihood and specification of a prior distribution for the parameters of the likelihood. We have shown that a Bayesian model for biofluid spectra, which exploits an informative parametric prior for the patterns of resonance generated by selected metabolites, can be used to deconvolve those resonances from a spectrum and to obtain explicit concentration estimates for the metabolites.

Simulations show that our MCMC algorithm usually identifies spectral resonance peaks precisely. Peak misalignment may occur when a target resonance for a multiplet of a template appears in the spectrum close to other stronger signals. The model may then encourage the template to align incorrectly with the stronger signals, even if they have the wrong shape. This is because the wavelet coefficients are heavily penalized in the prior but the parametric templates are not. Even when the model posterior concentrates around an incorrect deconvolution, the strong prior penalization on negative spectral signal means that posterior estimates of concentration can still be accurate, providing at least one multiplet for each metabolite deconvolves correctly. Concentration estimation, the main motivation for the modeling, is therefore quite robust to mis-assignment of spectral resonances.

It is worth noting that resonance mis-assignment is a problem for all methods, (including manual assignment by an expert; it is unavoidable for binning methods when peaks overlap) and our approach suggests two methods for resolving mistakes.  Firstly, signature templates corresponding to the compounds generating confounding signals can be added to the parametric component of the model (providing they are available). Secondly, the prior on the chemical shift parameter can be adjusted to fix the position of a misaligned multiplet.

Our approach yields improved concentration estimates. A comparison with a method for quantifying metabolites based on numerical integration shows the posterior mean estimates of the Bayesian model to be $6$ fold more accurate in quadratic error, even when exact multiplet locations are given to the numerical integration algorithm.

We are able to fit $M=20$ metabolites to an $n=3~000$ datapoint spectrum in about $60$ minutes on a $2.2\textrm{GHz}$ desktop machine using less than $0.5\textrm{GB}$ of RAM  when running the MCMC algorithm for $10 ~000$ iterations.  (If multiple chains are required to improve mixing, they will run in parallel on a multicore machine.)  The number of operations required by the MCMC algorithm is linear in $n$ and $M$.  The memory requirement is linear in the number of MCMC iterations and quadratic in $n$. This rate of computation can easily compete with the rate of acquisition of spectra by a typical NMR laboratory, effectively removing a major bottleneck in laboratory pipelines.

An accurate, automatic method for estimating metabolite concentrations from \textsuperscript{1}H-NMR spectra will assist many research projects in metabolomics. The field relies heavily on NMR for metabolite quantification and currently, even projects analyzing a few tens of spectra use numerical integration for estimation. Bayesian modeling should become increasingly useful as prior information on metabolite resonance patterns becomes more accurate and extensive. For example, with more detailed information our template model could be extended to deal systematically with the effects of interactions between multiplets. We plan to develop our model further and to release an efficient implementation of our methodology capable of simultaneously deconvolving the majority of metabolites assignable in the NMR spectra of complex biological mixtures.\footnote{BATMAN, an R package based on a C++ implementation of our methodology, is available from URL: http://www.ic.ac.uk/medicine/people/t.ebbels/}

\bibliographystyle{ECA_jasa}
\bibliography{bayes_nmr}

\end{document}